\documentclass[useAMS,usenatbib]{mnras}
\usepackage{graphicx,verbatim}
\usepackage[normalem]{ulem}
\usepackage{color,ulem}
\usepackage{newtxtext}
\usepackage[T1]{fontenc} 
\usepackage{amsmath,amssymb,bm}
\usepackage{cancel}

\def \be{\begin{equation}}
\def \ee{\end{equation}}
\newcommand       \ba           {\begin{eqnarray}}
\newcommand       \ea           {\end{eqnarray}}
\def \la{\langle}

\def \bea{\begin{eqnarray}}
\def \eea{\end{eqnarray}}

\def\bm#1{\mbox{\boldmath $#1$}}
\newcommand{\comments}[1]{}

\definecolor{webgreen}{rgb}{0,.5,0}
\definecolor{webbrown}{rgb}{.6,0,0}
\hypersetup{%
   colorlinks=true,%hyperfootnotes=false,%
   breaklinks=true,%
   plainpages=false, bookmarksnumbered, bookmarksopen=true,
   bookmarksopenlevel=1,%
   urlcolor=webbrown, linkcolor=webbrown, citecolor=webgreen,
   }

\title[Characterising MRI driven dynamo]{Characterising the Dynamo in a Radiatively Inefficient Accretion Flow}   

\author[ Dhang, Bendre, Sharma, \& Subramanian]
{Prasun Dhang$^{1,3,4}$,\thanks{E-mail:prasundhang@gmail.com}
Abhijit Bendre$^{2}$,\thanks{abbendre@gmail.com}
Prateek Sharma$^{1}$,\thanks{prateek@iisc.ac.in}
Kandaswamy Subramanian$^{2}$\thanks{kandu@iucaa.in}
\\
$^{1}$Department of Physics and Joint Astronomy Programme, Indian Institute of Science, Bangalore, INDIA 560012\\
$^2$ IUCAA, Post Bag 4, Ganeshkhind, Pune 411 007, INDIA\\
$^3$ Institute for Advanced Study, Tsinghua University, Beijing 100084, China\\
$^4$ Department of Astronomy, Tsinghua University, Beijing 100084, China
}

\begin{document}
\maketitle
\label{firstpage}
\begin{abstract}
We explore the MRI driven dynamo in a radiatively inefficient accretion flow (RIAF) using the mean field dynamo paradigm. 
Using singular value decomposition (SVD) we obtain the least squares fitting dynamo coefficients $\alpha$ and $\gamma$ by comparing the time series of the turbulent electromotive force and the mean magnetic field. Our study is the first one to show the poloidal distribution of these  dynamo coefficients in global accretion flow simulations. Surprisingly, we obtain a high value of the turbulent pumping coefficient $\gamma$ which transports the mean magnetic flux radially outward. This would have implications for the launching of magnetised jets which are produced efficiently in presence a large-scale poloidal magnetic field close to the compact object. We present a scenario of a truncated disc beyond the RIAF where a large scale dynamo-generated poloidal magnetic field can aid jet-launching close to the black hole. Magnitude of all the calculated coefficients decreases with radius. Meridional variations of $\alpha_{\phi \phi}$, responsible for 
toroidal to poloidal field conversion, is very similar to that found in shearing box simulations using the `test field' (TF) method. By estimating 
the relative importance of $\alpha$-effect and shear, we conclude that the
MRI driven large-scale dynamo, which operates at high latitudes 
beyond a disc scale height, is essentially of the $\alpha-\Omega$ type. 
\end{abstract}


% ...................................................................................................................Introduction.......................................................................................................................%
\section{Introduction}
Angular momentum transport in a completely ionized rotationally-supported accretion disc (such as in a black hole binary [BHB])
is supposed to be mediated by a weak field instability; namely, the Magneto-rotational instability (MRI; \citealt{Velikhov1959, Chandrasekhar1960a, Balbus1991}). Although linear MRI guarantees outward angular momentum transport, its saturation determines different accretion properties such as accretion rate and luminosity. With the increase in computational capabilities, it has been 
possible in last two decades to study saturation of the MRI using both local (shearing box) 
and global simulations with increasing resolution. Previous local (\citealt{Brandenburg1995, Hawley1996a, Davis2010, Gressel2015,Bhat2016}) and 
global (\citealt{Flock2012a, Hawley2013, Parkin2013a, Suzuki2014, Jiang2014b, Hogg2018a}) studies
show that an MRI driven dynamo can sustain magnetic fields  in saturation, overcoming the  dissipative effects (for a review see \cite{Blackman2015}). They also
 show that the MRI driven dynamo generates 
 large scale magnetic fields. Large scale magnetic fields are not only the necessary ingredient
to produce outflows/jets (\citealt{Blandford1977, Blandford1982}),  they are also important in determining the level of angular momentum transport (\citealt{Johansen2008a, Bai2013})
in accretion discs. 
 
 Till date, most studies focus on dynamo action
in the standard disc either using a local or a global approach. All the global (\citealt{Arlt2001b, Flock2012a, Hogg2018a}) and 
a few local shearing-box (\citealt{Brandenburg1995, Davis2010}) simulations use a very simple mean field closure 
(equation \ref{eq:simple_alpha}) to characterise the dynamo coefficients in shearing box simulations. A few studies (\citealt{Brandenburg2008f, Gressel2010, Gressel2015}) 
consider a more complicated closure encapsulating the 
anisotropic nature of MHD turbulence. Using state of the art test field (TF) method (\citealt{sch_07, Brandenburg2009d}), these studies determine turbulent dynamo coefficients for the  MRI driven dynamo.

In this paper, we wish to characterise the mean field dynamo in a  hot, optically thin, geometrically thick radiatively inefficient accretion flow (RIAF; \citealt{Narayan1994, Blandford1999, Narayan2000, Quataert2000, Yuan2012,Yuan2014}). We use the model `M-2P' described in \cite{Dhang2019} to calculate the dynamo coefficients.
Most of the previous studies determining the dynamo coefficients in local shearing box simulations use the TF method. We take an
alternate approach using singular value decomposition (SVD) method (\citealt{racin_11, simard_16, Bendre2019}). 
In this method, we essentially solve the problem of a least-square minimisation by fitting the time series of turbulent EMF with that of the 
mean magnetic fields. The advantage of SVD method is that we can post-process the simulation data,  while the TF method requires to solve additional equations for the passive fluctuating fields generated by large scale test fields.

It is thought that there is a close connection between the jet and the existence of a RIAF close to the black hole (\citealt{Esin1997, Fender1999a}).
Most of the previous studies investigating the disc-jet coupling assume the presence of a large-scale  magnetic field 
(\citealt{Tchekhovskoy2011, McKinney2012, Penna2013}). However, the source of
this coherent large-scale magnetic fields is still an open question. The two possible generation mechanisms
are (i) advection of field from the outer standard thin disc (\citealt{Bisnovatyi-Kogan1974a, Bisnovatyi-Kogan1976}) and (ii) in-situ generation of the field by dynamo action
(\citealt{Brandenburg1995, Hawley1996a}). In a classical turbulent thin disc ($H/R\ll1$), the advection timescale is comparable to the turbulent diffusion timescale (\citealt{Lubow1994, Cao2018}). However, a few studies propose that in the hot tenuous coronal region (where
radial velocity is comparatively higher compared to that in  the mid-plane) above and below the disc mid-plane, field can be dragged inward efficiently (\citealt{Guilet2012, Guilet2013}). There are some recent studies (\citealt{Hogg2018a, Liska2018a}), including ours (\citealt{Dhang2019}),
which investigate dynamo action in RIAFs. Large scale dynamo action is found to be weak in RIAFs and only confined beyond the disc scale height, within which a turbulent small-scale dynamo dominates. 

By using the SVD method, we obtain the distribution of turbulent dynamo coefficients in the poloidal ($r, \theta$) plane for a RIAF. We emphasise the previously unnoticed effect
of strong radial turbulent pumping (or what is known as the $\gamma$-effect), which transports large-scale magnetic fields radially outward. Presence of this $\gamma-$effect makes it harder for the large-scale magnetic field to be advected towards the black hole even in a weakly magnetized RIAF (like model `M-2P'). We propose a possible scenario for flux accumulation near the black hole in the truncated disc model (\citealt{Esin1997}) which is favourable for jet formation in the low/hard state of a BHB.

The paper is organised as follows. In section \ref{sect:method} we briefly discuss the details of the simulation which we analyse. The mean field formalism and the SVD method
are discussed in section \ref{sect:mfd_formalism}. 
We describe the results in section \ref{sect:results}. A discussion of 
the results and their astrophysical implications
is presented 
in section \ref{sect:discussion}. Finally, we summarise the key findings of the paper in section \ref{sect:summary}.

\section{The simulation details}
\label{sect:method}
We retrieve the dynamo coefficients for the MRI driven dynamo in a RIAF. 
We use the model `M-2P' described in \cite{Dhang2019}. To summarise, we 
solve the Newtonian ideal MHD equations of motion of the magnetised gas 
around a non-spinning black hole using the {\tt PLUTO} code 
(\citealt{Mignone2007}). Considering an ideal equation of state with 
adiabatic index $\Gamma=5/3$, we solve the equations in spherical 
coordinates ($r, \theta, \phi$) with the computational domain spanning 
over $r \in [4, 140]$, $\theta \in [0.02, \pi-0.02]$, $\phi \in [0, 2 
\pi]$ with a resolution $N_r \times N_{\theta} \times N_{\phi}=368 
\times 192 \times 512$. Most of the grids are employed in the region  
$r \in [4, 45]$, $\theta \in [\pi/3, 2\pi/3]$, $\phi \in [0, 2 \pi]$ 
where bulk of the accreting gas is present (for a detailed description 
see Table 1 and section 2.3 of \cite{Dhang2019}). 

In the Newtonian regime, to mimic the general relativistic effects close 
to a black hole, we use the pseudo-Newtonian potential 
(\citealt{Paczynsky1980}) $\Phi = GM/(r-2r_g)$, where $r_g = GM/c^2$ is 
the gravitational radius, $M$ and $c$ are the mass of the accreting black 
hole and the speed of light in vacuum respectively. In the code, we assume 
$GM=c=1$ to work in dimensionless units. As a result, all the length scales 
and velocities are expressed  in the units of $r_g$ and $c$ respectively. 
Unless stated otherwise, time scales are expressed in terms of the number 
of orbits a test particle would do at  the {\em inner most stable circular 
orbit} (ISCO), and is given by
\be
N_{\rm ISCO} = \frac{t_{\rm sim}}{T_{\rm ISCO}},
\ee
where the simulation time is $t_{\rm sim}$, the orbital period at ISCO, 
$T_{\rm ISCO}= 2 \pi r_{\rm ISCO}^{1/2}(r_{\rm ISCO}-2)= 61.56$ $r_g/c$ and we use $r_{\rm ISCO} = 6 r_g$ for a Schwarzschild black hole.

 We initialise a constant angular momentum torus (\citealt{Papaloizou1984}), embedded in a non-rotating, low-density hydrostatic medium. A poloidal magnetic field of plasma beta $\beta_{\rm ini}=890$ is initially threading the equilibrium torus and is parallel to the density contours.

For the extraction of dynamo coefficients, we use the time series of mean 
magnetic fields and EMFs in the quasi-steady state with a span $t_{\rm sim} \in [250,630]$ and data dumping interval $\Delta t=t_{i+1}-t_{i}=1.61$ $N_{\rm ISCO}$. Also, we only cover the radial range $r \in \left [10, 50 \right]$ $r_g$ which is roughly in a  statistically stationary state, given that the inflow equilibrium radius is $r_{\rm eq} = t_{\rm visc} v_r \approx 40$ $r_g$ (see section 5.7 of \cite{Dhang2019}). Here, $t_{\rm visc}$ and $v_r$ are the viscous time and radial velocity respectively.

\section{The Mean Field Dynamo}
\label{sect:mfd_formalism}
 To understand the evolution of large-scale magnetic fields in the 
simulations, we employ the standard mean-field dynamo formalism. Its mathematical formulation 
of relies upon splitting of the flow variables, 
velocity $ \bf{v}$ and magnetic field $\bf{B}$ as a sum 
of large scale or mean components (denoted by over-bars
, $\bar{\bf{v}}$ and $\bar{\bf{B}}$) and small-scale or 
fluctuating components (denoted by primes 
$\bf{v}^{\prime}$ and $\bf{B}^{ \prime }$),
where the 
mean is defined over a suitable averaging domain that
usually satisfies Reynolds rules. We thus define mean of a
quantity $q$ by integrating over the azimuthal domain as,
\be
\label{eq:mean_def}
\bar{q}(r,\theta,t) = \frac{1}{2 \pi} \int_{0}^{2 \pi} q(r,\theta,\phi,t) % d\phi.
\ee
and consequently the fluctuations as
\be
\label{eq:fluc_def}
q^{\prime}(r,\theta,\phi,t) = q(r,\theta,\phi,t) - \bar{q}(r,\theta,t).
\ee

Time evolution of the mean magnetic field is then governed 
by averaged induction equation,
 \be
  \label{eq:mean_B_evo}
 \frac{\partial \bar{\bf{B}}}{\partial t} = \nabla \times \left [ \bar{\bf{v}} \times \bar{\bf{B}} + \overline{\bf{v}^{\prime} \times \bf{B}^{\prime}} - \eta  \nabla \times \bar{\bf{B}} \right]
 \ee
where $\eta$ is the microscopic diffusivity. The terms 
within the bracket on the right-hand side describe
the effects of mean and fluctuating fields and flows on its evolution. 
The effect of turbulence on the mean field 
evolution is captured through the mean electromotive force 
(EMF),
 \be
 \label{eq:mean_emf}
 \bar{\mathcal{E}} = \overline{\bf{v}^{\prime} \times \bf{B}^{\prime}}.
 \ee 
Motivated by Second Order Correlation Approximation (SOCA) 
\citep{Mof78,KR80},
the EMF is modelled as linear function of mean field and 
its covariant derivatives ignoring any higher order
derivatives. In a locally Cartesian co-ordinate system we have, 
\be
\label{eq:mean_emf_soca}
\bar{\mathcal{E}}_i = a_{ij}\bar{B}_j + b_{ijk}\frac{\partial\bar{B}_{j}}{\partial x_k}.
\ee
Here the tensorial 
coefficients $a_{ij}$ and $b_{ijk}$ are the functions of 
statistical correlation among fluctuating parts of velocity 
and magnetic fields. Further, $\bf{a}$ is a rank two tensor which acts 
as a source term in equation \ref{eq:mean_B_evo}, and $\bf{b}$, 
a rank three tensor is mainly responsible for diffusion of 
mean magnetic field. By further decomposing these tensors 
into their symmetric and anti-symmetric parts, equation \ref{eq:mean_emf_soca} can be rewritten as,
\bea
\label{eq:emf_alpha_gamma}
\bar{\mathcal{E}_i} = \alpha_{ij} \bar{B}_j 
+ \left(\bm{\gamma} \times \bar{\bf{B}}\right)_{i}
-\eta_{ij} (\nabla \times \bar{\bf B})_{j} \nonumber\\
-\left[\bf{\Delta} \times (\nabla \times \bar{\bf{B}})\right]_{i}
-\kappa_{ijk}\frac{\partial \bar{\rm B}_{j}}{\partial {x_{k}}}
\eea
\cite[See eg. ][and references therein. ]{sch_07}
Here the dynamo coefficients $\alpha, \gamma, \eta, 
\Delta$ and $\kappa$ describe the various turbulence 
properties, eg.
\be
\label{eq:dynamo_coefficients}
\alpha_{ij} = \frac{1}{2}(a_{ij} + a_{ji}), ~{\rm and}~ \gamma_k = \frac{1}{2}\epsilon_{kij}a_{ij}
\ee
describe the classical $\alpha$-effect and turbulent pumping 
respectively. 
It can be explicitly seen from equation \ref{eq:emf_alpha_gamma} and equation \ref{eq:mean_B_evo}
that $\bm{\gamma}$ adds to the mean velocity $\bar{\mathbf{v}}$ in advecting the mean magnetic field.
The tensor $\eta_{ij}$ represents the diagonal
and off-diagonal parts of anisotropic turbulent diffusivity, while the coefficient $\bm{\Delta}$ encapsulates a dynamo generating term first identified by \citet{Radler}. In this analysis however we have neglected the contribution of 
tensor $\bf{b}$, since it had no significant impact on the 
determination of EMF.

\subsection{Estimating Dynamo Coefficients Using the SVD Method}
The dynamo coefficients in equation \ref{eq:emf_alpha_gamma} express the effects of MHD turbulence on the evolution of the mean field. It is therefore of 
interest to determine these coefficients for the direct 
simulations we have performed. To explicitly state the 
problem we first express the $i^{th}$ component EMF using 
equation \ref{eq:mean_emf_soca} and ignore the higher order 
terms involving tensor $\mathbf{b}$ in the expansion 
\citep[similar to][]{racin_11} as, 
\be
\bar{\mathcal{E}}^{M}_i = {a}_{ij} \bar{B}_j,
\label{eq:emf_sph_pol}
\ee
where $i,j \in (r, \theta, \phi)$, and the pseudo-tensor 
$a_{ij}$ relates to the direct dynamo coefficients 
$\alpha_{ij}$  and $\gamma_i$ in
equation \ref{eq:dynamo_coefficients} through following 
relations,
\bea
\label{eq:alphas}
\alpha_{rr}             &=& {a}_{rr},\nonumber\\
\alpha_{\theta \theta}  &=& {a}_{\theta \theta},\nonumber\\
\alpha_{\phi\phi}       &=& {a}_{\phi \phi},\nonumber\\
\alpha_{r\theta}        =& \alpha_{\theta r} = &\frac{1}{2}\left(
                                {a}_{r \theta} + {a}_{\theta r}\right) \nonumber\\
\alpha_{r\phi}          =& \alpha_{\phi r} =& \frac{1}{2}
                                \left({a}_{r \phi} + {a}_{\phi r} 
                                \right),\nonumber\\
\alpha_{\theta\phi}     =& \alpha_{\phi\theta} = &\frac{1}{2}
                                \left({a}_{\theta \phi} + {a}_{\phi \theta} 
                                \right).
\eea
\bea
\label{eq:gammas}
\gamma_r        &=&   \frac{1}{2}
                    \left(
                        {a}_{\theta\phi}
                    -   {a}_{\phi\theta}
                    \right),\nonumber\\    
\gamma_\theta   &=&   \frac{1}{2}
                    \left(
                        {a}_{\phi r}
                    -   {a}_{r \phi}
                    \right),\nonumber\\   
\gamma_\phi     &= &  \frac{1}{2}
                    \left(
                        {a}_{r\theta}
                    -   {a}_{\theta r}
                    \right).   
\eea
The problem then is one of determining ${a}_{ij}$ and from the data of 
DNS, at each point in $ \left(r,\theta\right)$. Note here 
that we have ignored the contribution of tensor $\mathbf{
b}$ as it was found to have no statistically significant 
effect on the determination of EMF, as discussed in the 
later sections. The expressions for the complete set of 
dynamo coefficients is given in Appendix 
\ref{app_all_coefficients} \citep[see also][ and references therein]{sch_07}.
Due to insufficient number of equations than are required to 
have a unique solution to 
equation \ref{eq:emf_sph_pol}, we treat this problem as the one 
of time series analysis. In particular, we adopt the standard 
singular value decomposition method to minimise the 
least-squares of the time series formed from the residuals of 
equation \ref{eq:emf_sph_pol} at each $ \left( r, \theta \right)$ 
point. We have adopted this method from \citet{racin_11} and 
\citet{simard_16} and the details are discussed below. 

At each point in $ \left( r, \theta \right) $ 
plane, the time series of EMF components $\bar{\mathcal{E}
}_i  $ and mean magnetic field $\bar{B}_i$ are 
extracted from the DNS. Elements of these time series are 
treated as independent data points, and used to determine 
the values of ${a}_{ij}$ at $\left(r,\theta\right)$. 
This is achieved by determining the set of coefficients 
that minimise the least-square sums of following residual 
vector components;
\be
%R_i = \bar{\mathcal{E}}_i - \bar{\mathcal{E}}^{M}_{i}
R_i = \bar{\mathcal{E}}_i - {a}_{ij} \bar{B}_j,
\label{eq:residual_vector}
\ee
by using the SVD scheme.

We first construct the `design matrix', $\mathcal{A}$ at  
each fixed $\left(r,\theta\right)$ point which we define as,
\bea
\mathcal{A} = \begin{pmatrix}
\bar{B}_r\left(t_1\right)& \bar{B}_\theta\left(t_1\right)&\bar{B}_\phi\left(t_1\right)\\[5pt]
\bar{B}_r\left(t_2\right)& \bar{B}_\theta\left(t_2\right)&\bar{B}_\phi\left(t_2\right)\\
\vdots                              & \vdots                                    &\vdots \\
\bar{B}_r\left(t_N\right)& \bar{B}_\theta\left(t_N\right)&\bar{B}_\phi\left(t_N\right)\\
\end{pmatrix}
\label{eq:design_matrix}
\eea
$\mathcal{A}$ is determined 
directly from the DNS data. The number of 
rows $N$, indicates the length of these time series, $\left(t_1,
t_2,...,t_N\right)$. Similarly we define the data matrix 
$\mathcal{Y}$ and the parameter matrix $\mathcal{X}$ (also at 
$(r, \theta)$) as,
\bea
\mathcal{Y} =      \begin{pmatrix}
\bar{\mathcal{E}}_r\left(t_1\right) & \bar{\mathcal{E}}_\theta\left(t_1\right)  & \bar{\mathcal{E}}_\phi\left(t_1\right)  \\[5pt]
\bar{\mathcal{E}}_r\left(t_2\right) & \bar{\mathcal{E}}_\theta\left(t_2\right)  & \bar{\mathcal{E}}_\phi\left(t_2\right)\\
\vdots                              & \vdots                                    &\vdots \\
\bar{\mathcal{E}}_r\left(t_N\right) & \bar{\mathcal{E}}_\theta\left(t_N\right) & \bar{\mathcal{E}}_\phi\left(t_N\right) 
\end{pmatrix},
\label{eq:data_matrix}
\eea
and 
\bea
\mathcal{X} =      \begin{pmatrix}
{a}_{r r}         &{a}_{\theta r}      &{a}_{\phi r}\\[5pt]
{a}_{r \theta}    &{a}_{\theta \theta} &{a}_{\phi \theta}\\[5pt]
{a}_{r \phi}		&{a}_{\theta \phi}    &{a}_{\phi \phi}
\end{pmatrix}.
\label{eq:parameter_matrix}
\eea
Noting that matrices $\mathcal{X}, \mathcal{Y}$ and $\mathcal{A}$ 
are of dimensions $3\times 3$, $N\times3$ and $N\times 3$ 
respectively, equation \ref{eq:emf_sph_pol} can be written as, 
\bea
\mathcal{Y}\left(r,\theta\right) = \mathcal{A}\left(r,\theta\right) \mathcal{X}\left(r,\theta\right) + \hat{\mathcal{N}}\left(r,\theta\right).
\label{eq:emf_matrix_eqn}
\eea
The additional term $\Hat{\mathcal{N}}$ on the right hand side  
is the noise matrix with same dimensions as $\mathcal{Y}$. 
Columns of $\Hat{\mathcal{N}}$ represent the level of additive 
noise in the components of $\bar{\mathcal{E}}$ while the rows
represent the additive noise in the respective EMF component 
at a fixed time. With this arrangement equation \ref{eq:emf_matrix_eqn} 
now represents an over-determined system of equations wherein 
the constancy of the elements of $\mathcal{X}$ is implicitly 
assumed for $t_1 < t < t_N$. Such an assumption appears reasonable
in the saturated steady state of the dynamo. The least squared solution ($
\Hat{\mathcal{X}}$) to this system is found by seeking to 
minimise the following two norm for $i^{th}$ columns of $
\mathcal{Y}$ and $\mathcal{X}$,
\bea
\chi_i^2\left(r,\theta\right) = \frac{1}{N}\sum_{n=1}^{N}\left[\frac{ \left(\mathbf{y}_i\left(r,\theta,t_n\right) - \mathcal{A}\mathbf{x}_i\left(r,\theta,t_n\right)\right)^\top }{\sigma_i}\right]^2.
\label{eq:two_norm}
\eea
Here the vectors $\mathbf{y}_i$ and $\mathbf{x}_i$ are the $i^{
th } $ columns of $\mathcal{Y}$ and $\mathcal{X}$ respectively. 
While $\sigma_i$ is the variance associated with $i^{th}$ column
of the noise matrix $\Hat{\mathcal{N}}$ (which we will determine
from the best fit $\Hat{\mathbf{x}}_i$ itself below). The method that we use
here to seek the least square solution $\Hat{\mathcal{X}}$, relies
upon the unique SVD decomposition of $\mathcal{A}$;
\bea
\mathcal{A} = \mathbf{U}\mathbf{w}\mathbf{V}^T
\label{eq:svd_decomp}
\eea
where the matrices $\mathbf{U}$ and $\mathbf{V}$ are of dimensions
$N\times3$ and $3\times3$ respectively and are orthonormal, while
the singular matrix $\mathbf{w}$ is $3\times3$ diagonal matrix. 
For this decomposition we follow the algorithm described in
\cite{recipies}. It is to be noted here that matrices $\mathbf{A}$, 
$\mathbf{U}$, $\mathbf{V}$ and $\mathbf{w}$ are functions of $r$ 
and $\theta$. We avoid 
the explicit mention of this fact 
hereafter, unless required. $\Hat{\mathcal{X}}$ can then be 
expressed in terms of $\mathbf{U}$, $\mathbf{V}$ and $\mathbf{w}$ 
simply as \citep[similar to ][]{mandel_82}, 
\bea
 \Hat{\mathbf{x}}_j = \mathbf{V}\mathbf{w}^{-1}\mathbf{U}^{T}\mathbf{y}_j.
\label{eq:least_sq_soln}
\eea
We recall that here the index `$j$' denotes the $j$'th column of either
$\Hat{\mathcal{X}}$ or $\mathcal{Y}$.
Variances associated with each element of $\Hat{
\mathbf{x}}_j$ do depend upon the column index $j$ (through the 
$\sigma_j$ in equation \ref{eq:two_norm}) as,
\bea
{\rm Var}\left(\left[\Hat{\mathbf{x}}_j\right]_i\right) = 
\sum_l\left[\frac{\mathbf{V}_{il}}{\mathbf{w}_{ll}}\right]^2 \sigma_j^2,
\label{eq:var}
\eea
where the noise component $\sigma_j$ is calculated from
the SVD fit simply as,
\bea
\sigma_j^2 = \frac{1}{N}\left(\mathbf{y}_j-\mathcal{A}\Hat{\mathbf{x}}_j\right)^\top\left(\mathbf{y}_j-\mathcal{A}\Hat{\mathbf{x}}_j\right).
\label{eq:sig}
\eea

\section{Results}
\label{sect:results}
Saturation of the MRI in the non-linear regime maintaining a finite
amplitude for the magnetic field indicates an underlying dynamo action. A distinguishing feature of the dynamo action in a geometrically 
thick RIAF is the presence of an intermittent dynamo cycle
(\citealt{Hogg2018a, Dhang2019, Liska2018a}). Irregularity in the dynamo cycle can 
readily be explained using the mean field dynamo theory (for a review see \citealt{Brandenburg2005}). Slightly sub-Keplerian nature of the angular 
velocity leads to the intermittency (\citealt{Nauman2015, Gressel2015, Dhang2019}).

 \subsection{Butterfly diagrams for the mean fields}
 \begin{figure}
    %\centering
    \includegraphics[scale=0.39]{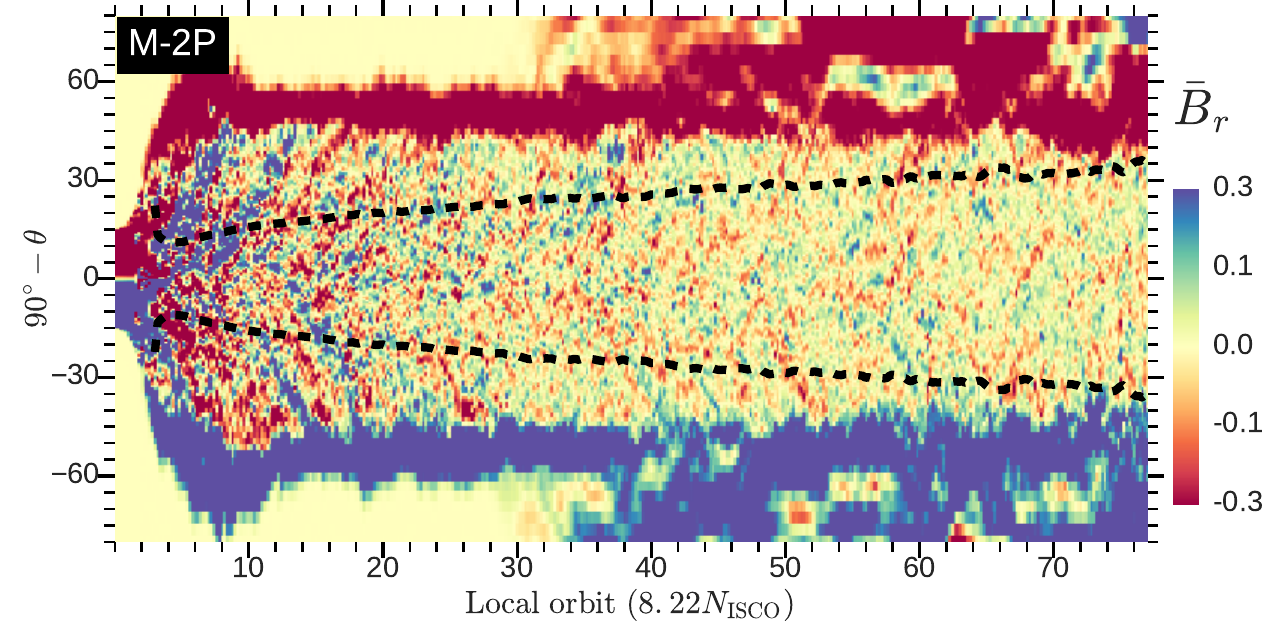}
    \includegraphics[scale=0.39]{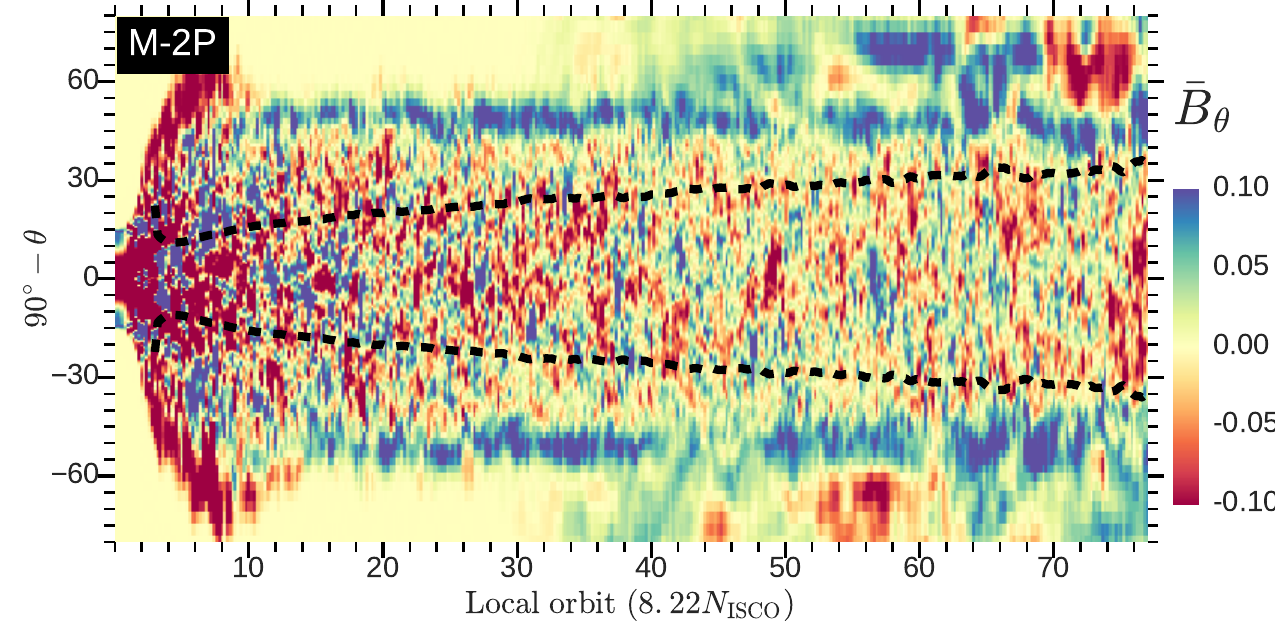}
    \includegraphics[scale=0.39]{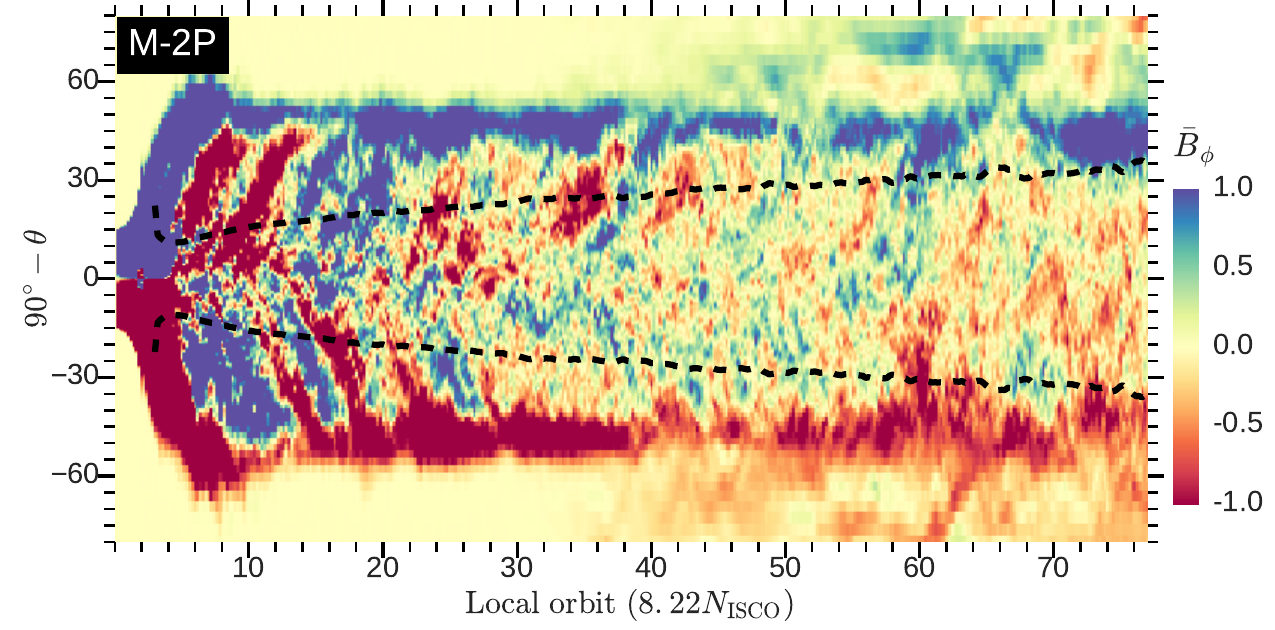}
 \caption{Spatio-temporal variation of radial (top panel), meridional (middle panel) and toroidal (bottom panel) components of mean magnetic fields at a radius $r= 20$. Time is expressed in units of local orbit at $r=20$. Black dashed lines indicate one scale-height in the northern and southern hemispheres.  Mean fields are stronger at larger latitudes on both the hemispheres, where large scale dynamo operates. Equatorial regions, where fluctuation dynamo dominates, have smaller mean fields.}
   \label{fig:b_mean}
 \end{figure}
 
 Fig \ref{fig:b_mean} shows the variation of the mean magnetic fields with latitude $\theta_l=90^{\circ}-\theta$ and time (popularly known as the `Butterfly diagram')
at a radius $r=20$ for a geometrically thick ($H/R \sim 0.4$) RIAF (model `M-2P' in \citealt{Dhang2019}). Top, middle and bottom panels of Fig. \ref{fig:b_mean} show the
 spatio-temporal variation of radial ($\bar{B}_r$), meridional ($\bar{B}_{\theta}$) and toroidal ($\bar{B}_{\phi}$) mean magnetic fields respectively. While $\bar{B}_{\phi}$ and $\bar{B}_r$ are the strongest components, strength of $\bar{B}_{\theta}$ is much smaller.
 It is also evident from Fig. \ref{fig:b_mean} that mean fields are stronger at higher latitudes, where stratification becomes important and large scale dynamo 
operates. On the other hand, at lower latitudes around the mid-plane, fluctuation dynamo dominates and mean fields are vanishingly small. It should also be mentioned that the contribution to the Maxwell stress from the mean fields are larger  where mean fields are predominant, while close to the mid-plane the Maxwell stress is mainly due to the correlation between the fluctuating components. For a 
detailed discussion on the two dynamo mechanisms in RIAFs see Fig. 26 and Section 7.4 in \cite{Dhang2019}. 
 
 \subsection{Spatial variation of the dynamo coefficients}
 \label{sect:dynamo_cof_spatial}
In an accretion flow, due to the presence of a strong shear, poloidal fields ($B_r$ and $B_{\theta}$) are readily converted into toroidal fields ($B_{\phi}$). 
The generation of poloidal fields is mainly attributed to the twisting of toroidal magnetic fields by  helical turbulence, known as the classical $\alpha$- effect. 
A simple mean field closure (equation \ref{eq:simple_alpha}) neglecting $\gamma$, turbulent diffusion and off-diagonal terms of $\alpha_{ij}$ fails to provide an ordered $\alpha$ coefficient for the dynamo action in RIAFs (see Fig. 25 of \cite{Dhang2019}). In the current work, we use a more general closure (equation \ref{eq:emf_sph_pol}) for mean fields and EMFs. 

\subsubsection{Meridional variation}
\label{sect:dynamo_cof_meridional}
\begin{figure*}
    %\centering
    \includegraphics[scale=0.55]{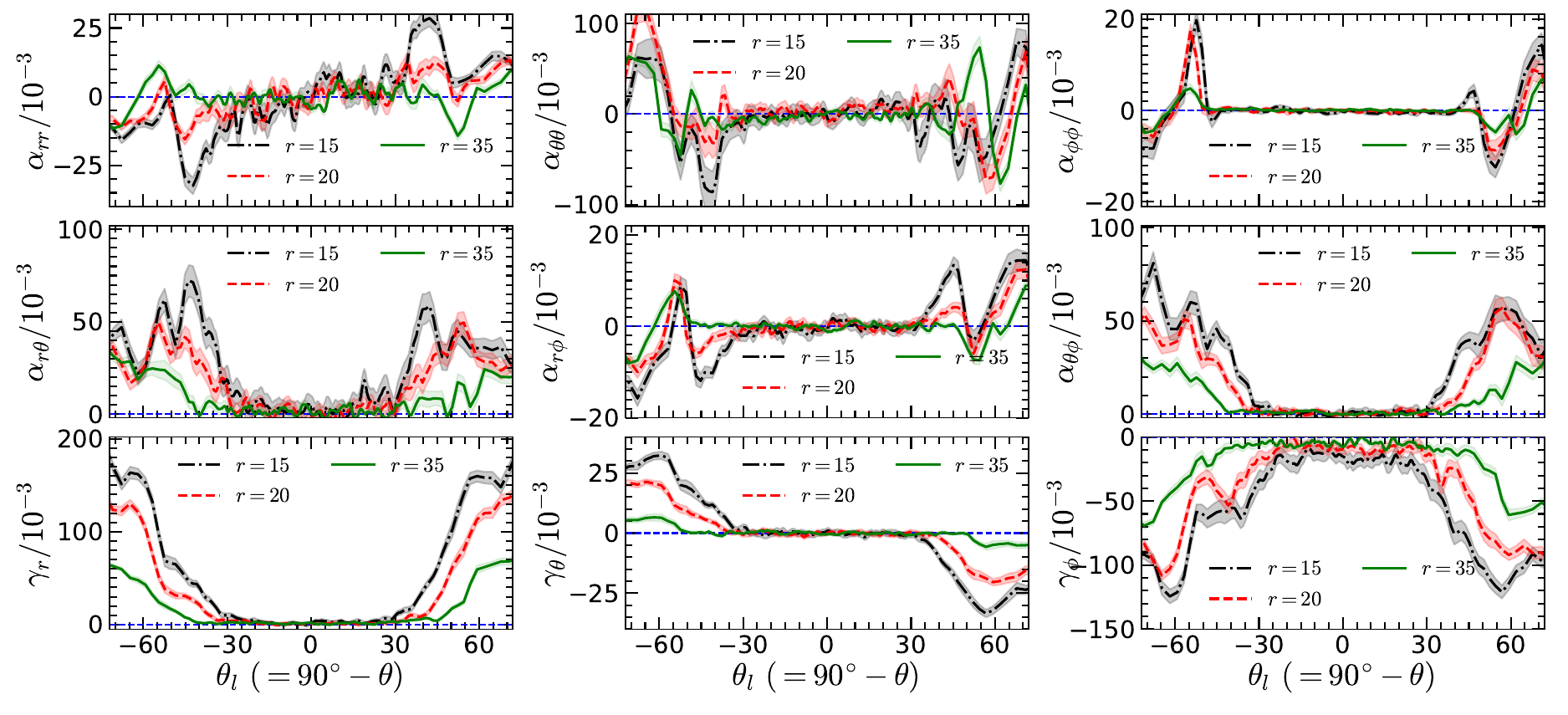}
 \caption{Meridional variation of dynamo coefficients at different radii $r=15, 20, 35$. Coloured region shows the corresponding $1 \sigma$ region of significance. Blue dashed horizontal line denotes zero of dynamo coefficients.}
   \label{fig:alpha_vertical}
 \end{figure*}

Fig. \ref{fig:alpha_vertical} shows the meridional variation of the nine dynamo 
coefficients at three different radii $r=15, 20, 35$ recovered using the SVD method. Top three panels in Fig. \ref{fig:alpha_vertical} show the variation of diagonal components of
$\alpha_{ij}$ with latitude $\theta_l=90^{\circ}-\theta$. While $\alpha_{rr}$ and $\alpha_{\phi \phi}$ are ordered and antisymmetric about the
mid-plane, $\alpha_{\theta \theta}$ shows less coherent structure and is symmetric about the mid-plane. Out of these three diagonal components, 
the most important one for the large-scale dynamo 
is the $\alpha_{\phi \phi}$, which is responsible for the production of poloidal fields $B_r$ and $B_{\theta}$ out of the toroidal field $B_{\phi}$. In the northern hemisphere (NH)
$\alpha_{\phi \phi}$ is negative at lower latitude and becomes positive at higher latitude. An opposite trend is
seen in southern hemisphere (SH). This antisymmetric behaviour of $\alpha_{\phi \phi}$ is expected from the naive picture of helical turbulence, 
where Coriolis forces break parity and the symmetry 
about the equator. The pattern in meridional variation of $\alpha_{\phi \phi}$ roughly matches with those found in local shearing box simulations (\citealt{Brandenburg2008f, Gressel2010}). However, the result differs from some of the previous global simulations (\citealt{Arlt2001b,Flock2012a}). We will discuss the sign of $\alpha_{\phi \phi}$ in different studies of MRI driven dynamo in detail in section \ref{sect:compare_works}.

Middle and bottom three panels in Fig. \ref{fig:alpha_vertical} show the meridional variation of symmetric ($\alpha_{r \theta}$, $\alpha_{r \phi}$, $\alpha_{\theta \phi}$) and antisymmetric ($\gamma_r$, $\gamma_{\theta}$, $\gamma_{\phi}$) parts of the off-diagonal components of $\bf{a}$-tensor. While $\alpha_{r \phi}$ and its 
corresponding anti-symmetric component $\gamma_{\theta}$ are anti-symmetric about the mid-plane, other components
show a symmetric behaviour. Among the non-diagonal components, the $\gamma$'s, which represent
 turbulent pumping, are of particular interest. The coefficients $\gamma$ represent the transport of the mean fields from a turbulent region to a more laminar region - a phenomenon of turbulent diamagnetism. Positive sign of radial component $\gamma_r$ at all latitudes implies a transport of mean fields from more turbulent region (close to the BH) to a comparatively less turbulent region (away from the BH). Similarly, the negative (positive) sign of $\gamma_{\theta}$ in the NH (SH) describes the pumping of mean fields from turbulent equatorial region to the more laminar coronal region. The coefficient $\gamma_{\phi}$ arises because of the non-alignment of angular velocity and gradient in total (magnetic + kinetic) specific turbulent energy (see equation 10.59 in \cite{Brandenburg2005}).

\subsubsection{Radial variation}
\label{sect:dynamo_cof_radial}
 \begin{figure*} 
    %\centering
    \includegraphics[scale=0.55]{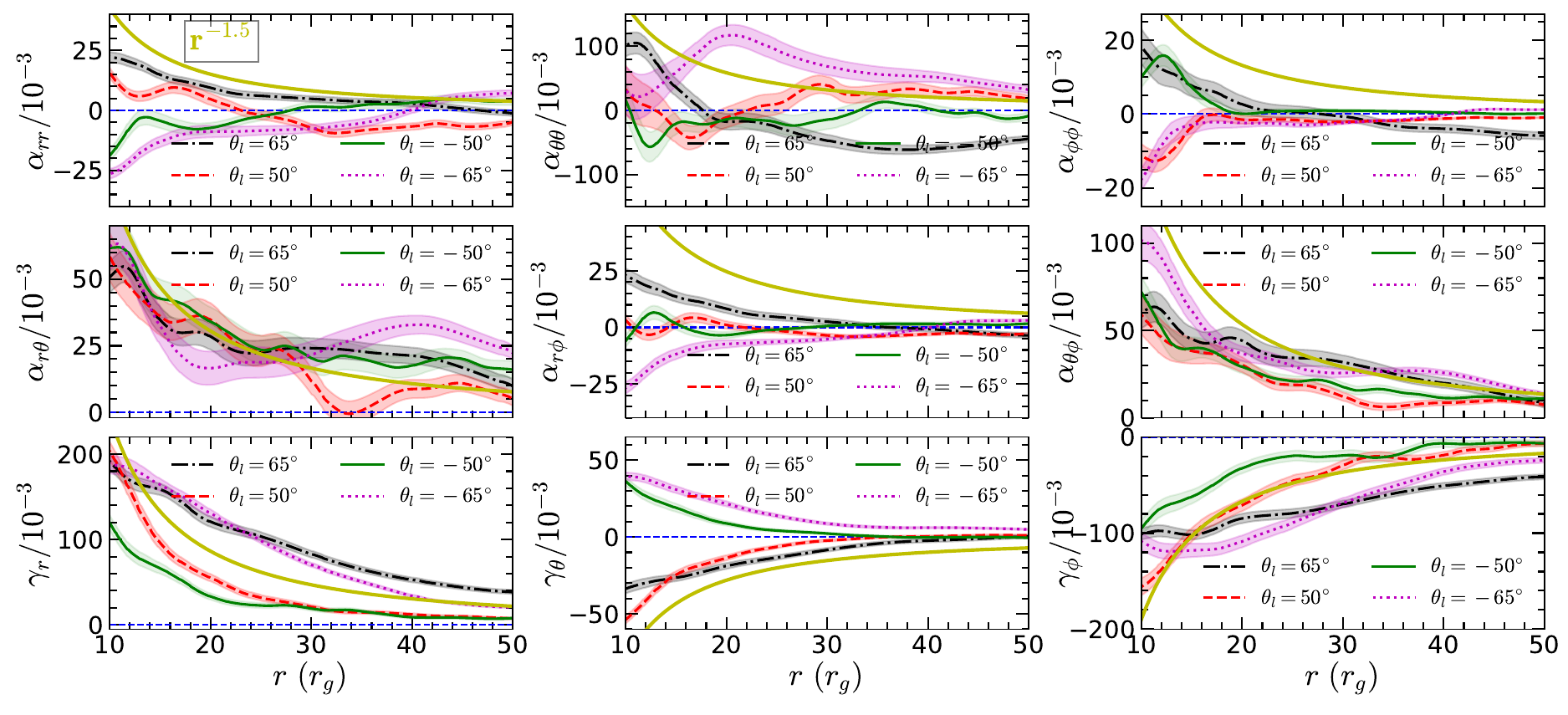}
 \caption{Variation of dynamo coefficients with  radius at four different latitudes $\theta_l=-65^{\circ}, -50^{\circ}, 50^{\circ}, 65^{\circ}$ in NH and SH. Corresponding $1 \sigma$ region of significance is shown by the coloured error band. Blue dashed horizontal line denotes zero of dynamo coefficients. A power law profile of $\propto r^{-1.5}$ is shown by the yellow solid line. }
   \label{fig:alpha_radial}
 \end{figure*}
 In Fig. \ref{fig:alpha_vertical}, we see a clear pattern, smaller the radius, larger the value of the 
 dynamo coefficients. To get a better picture of this behaviour, 
 we look at variation of the dynamo coefficients 
 at high latitudes (where large-scale dynamo dominates) with radius in Fig. \ref{fig:alpha_radial}. We take four meridional cuts - two in the NH (with $\theta_l=50^{\circ}, 65^{\circ}$), and two in the SH ($\theta_l=-50^{\circ}, -65^{\circ}$). Selection of such meridional cuts helps us comprehend the radial variation as well the symmetry of the dynamo coefficients about the mid-plane.   
 
 Top, middle and bottom panels of Fig. \ref{fig:alpha_radial} show the radial variation of diagonal, symmetric and antisymmetric
 components of the $\bf{a}$ tensor respectively. Among the diagonal components, both $\alpha_{rr}$ and $\alpha_{\phi \phi}$ show a coherent
 antisymmetric behaviour at all radii (e.g. compare black and magenta lines representing dynamo coefficients at $\theta_l=+65^{\circ}$
 and $\theta_l=-65^{\circ}$ respectively). The symmetric nature of $\alpha_{\theta \theta}$ as we describe in section
 \ref{sect:dynamo_cof_meridional} is not prevalent at all radii. Like $\alpha_{rr}$ and $\alpha_{\phi \phi}$, the symmetric/antisymmetric
 behaviour of the off-diagonal components are quite robust. The coefficients $\alpha_{r \phi}$ and $\gamma_{\theta}$ are anti-symmetric about the
 mid-plane at all radii. Rest of the off-diagonal components preserve symmetry about mid-plane which we discuss in section \ref{sect:dynamo_cof_meridional}. 
 
 Till now, in this sub-section, we discussed how the spatial symmetry of the dynamo coefficients prevails at all radii. However, the most 
 interesting result of the our SVD analysis in this sub-section is that 
 many of the calculated dynamo coefficients roughly follow a power-law of $\propto r^{-1.5}$. To guide reader, we draw a yellow solid line following a power-law $\propto r^{-1.5}$ in each panel of Fig. \ref{fig:alpha_radial}. In our simulations of the RIAF, we have $\Omega \sim r^{-1.7}$ and it is thus interesting that dynamo coefficients can have the similar radial dependence.
 Moreover, the coefficient crucial for the generation of poloidal field from toroidal one, $\alpha_{\phi \phi}$, is also negative 
 in the NH and tends to change sign again further north. It would be important to understand these interesting features of $\alpha_{\phi \phi}$ from 
 more basic theory in the future.

\subsubsection{Variation in the poloidal plane- a composite picture}
 \label{sect:dynamo_cof_poloidal}
\begin{figure*}
    %\centering
    \includegraphics[scale=0.45]{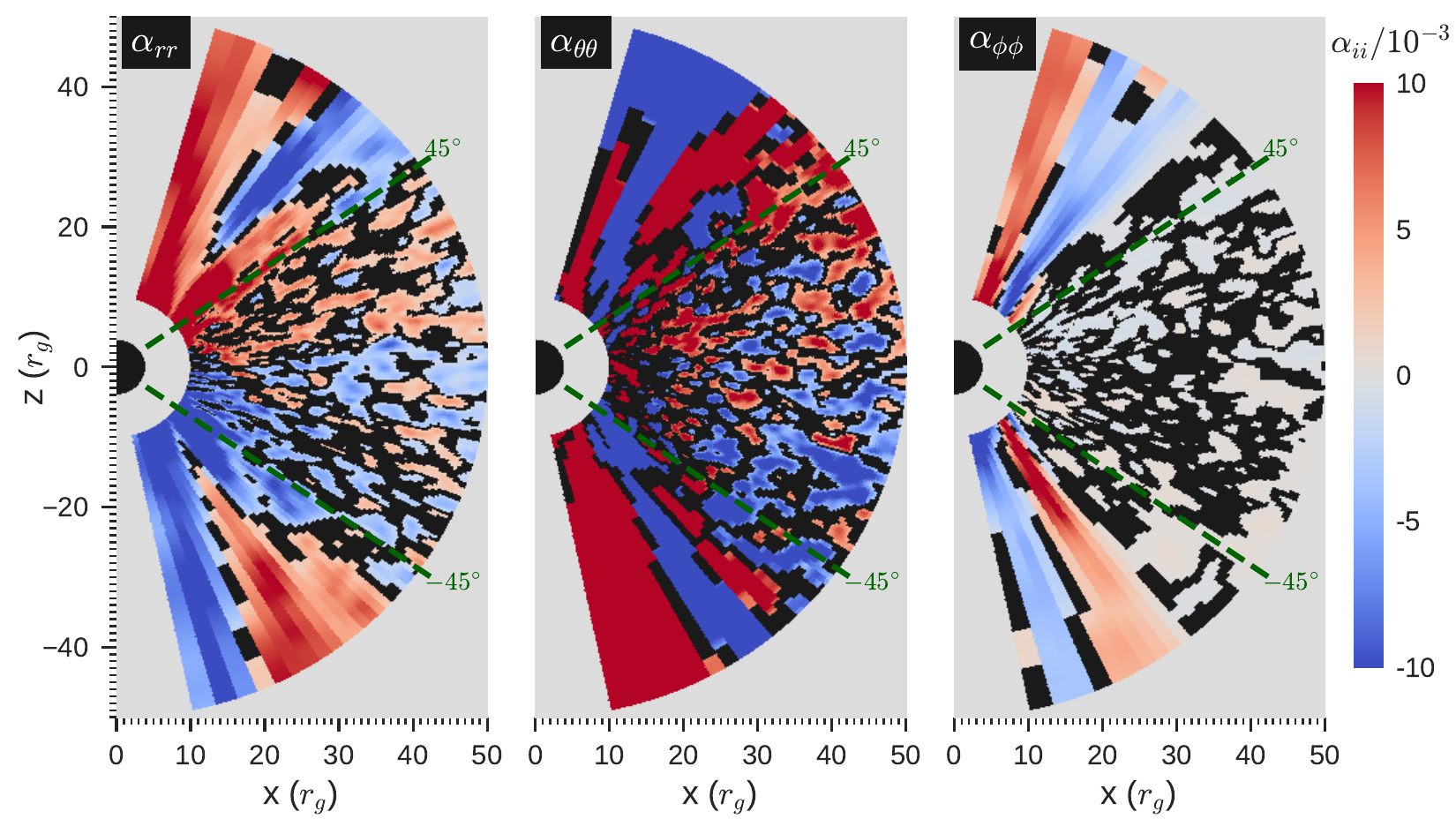}
    \includegraphics[scale=0.45]{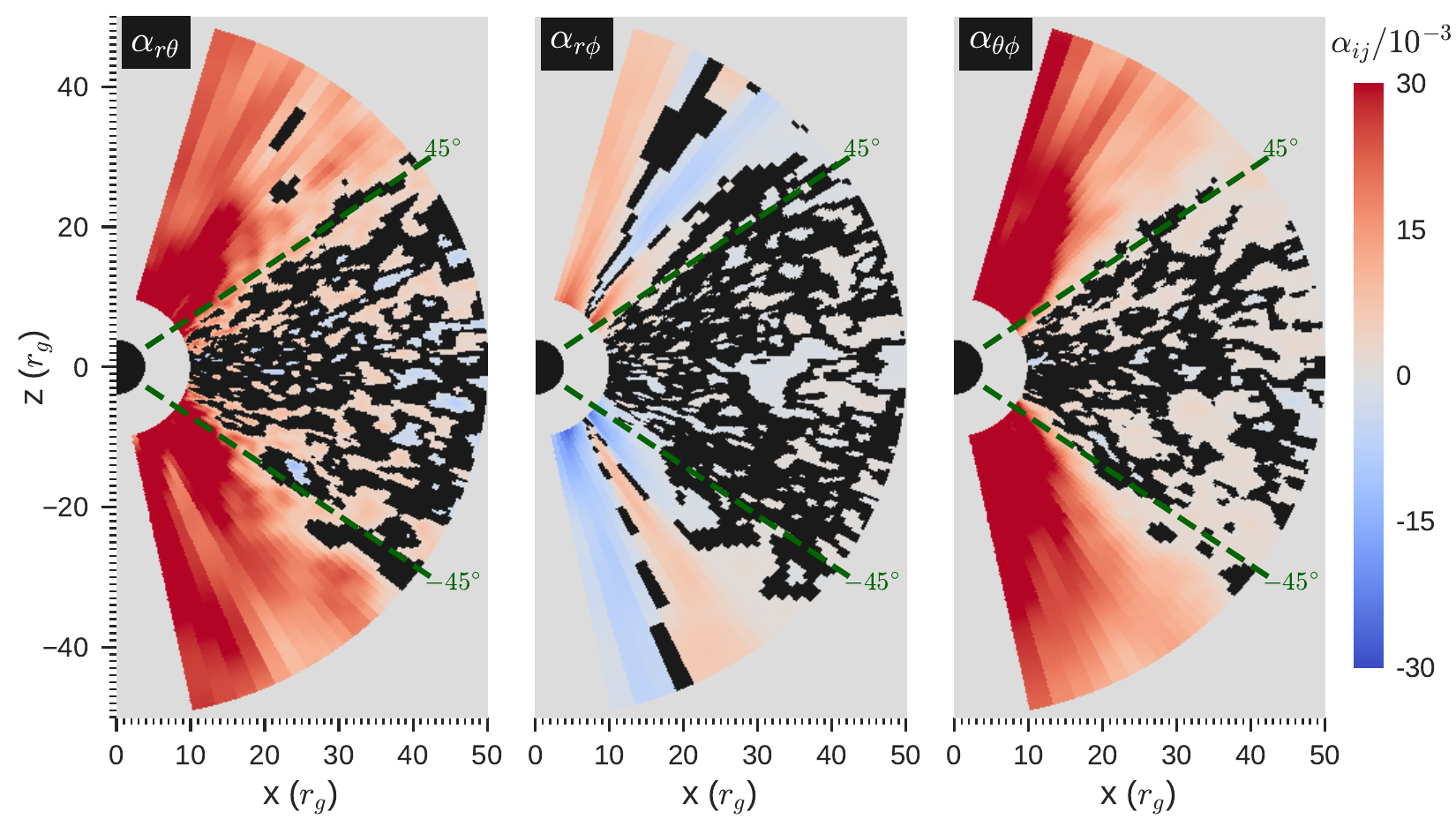}
    \includegraphics[scale=0.45]{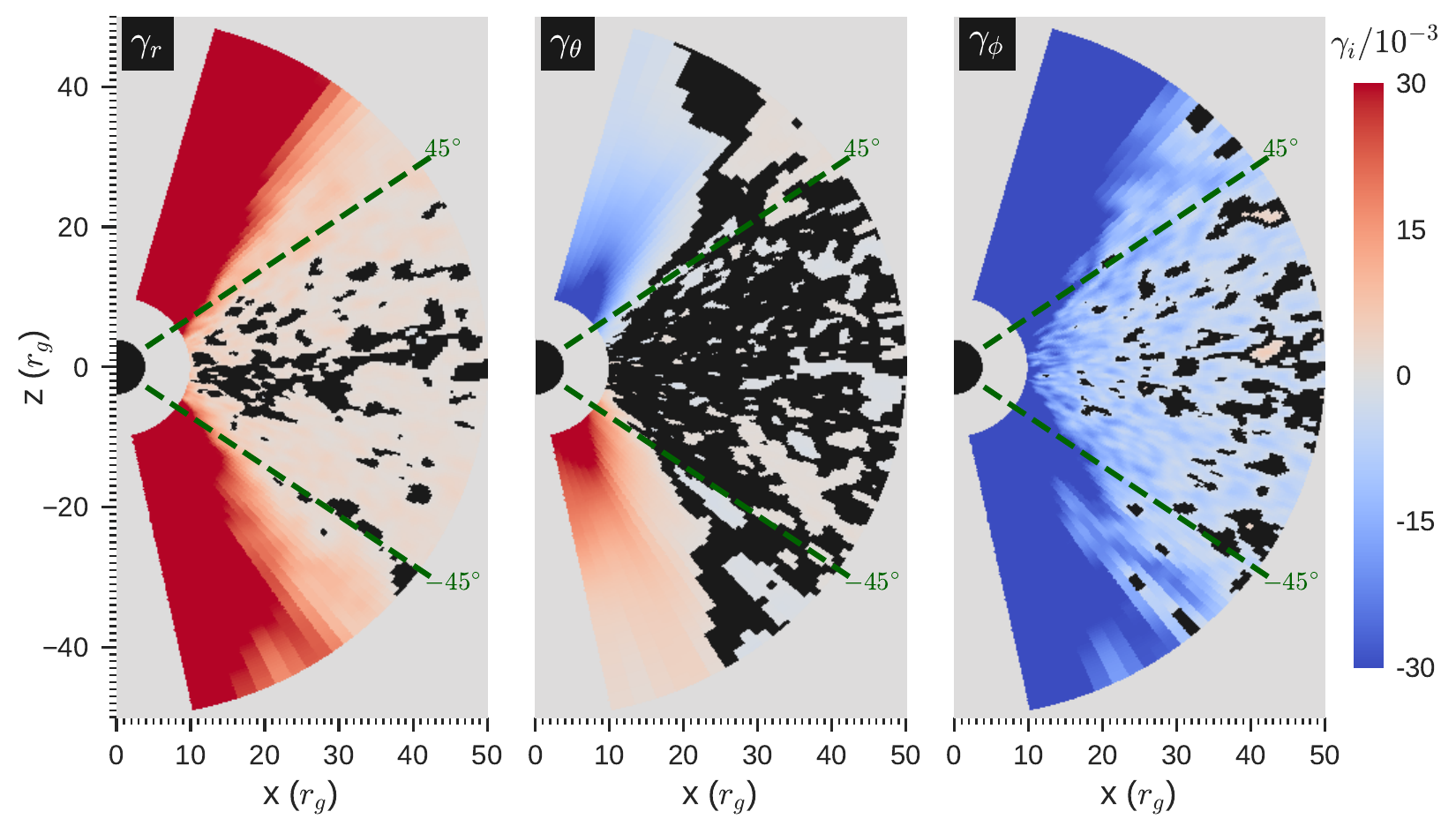}
 \caption{Poloidal distribution of different dynamo coefficients obtained by the SVD method using time series of the mean fields and EMFs in the quasi-steady state ($t_{\rm sim} [250, 630]$). The regions where the dynamo coefficients are smaller than  one standard deviation, 
a black mask is applied.  Green dashed lines on the NH and SH denote the latitudes $\theta_l=45^{\circ}$ and $\theta_l=-45^{\circ}$ respectively. Two of the diagonal ($\alpha_{rr}$, $\alpha_{\phi \phi}$), and off-diagonal ($\alpha_{r \phi}$ and $\gamma_{\theta}$) components are anti-symmetric about the mid-plane ($\theta_l=0^{\circ}$). Rest of the off-diagonal coefficients ($\alpha_{r \theta}$, $\alpha_{\theta \phi}$, $\gamma_r$ and $\gamma_{\phi}$) show symmetric behaviour about the mid-plane; $\alpha_{\theta \theta}$ is mildly symmetric about the mid-plane. Symmetry/antisymmetry of the dynamo coefficients about the mid-plane is not an artifact of the SVD method as 
the method treats each point in space independently.}
   \label{fig:alpha_poloidal}
 \end{figure*}
 
 In the previous sub-sections, in order to study spatial dependence, we determined  
 the meridional and radial variations of dynamo coefficients at the fixed radii and latitudes respectively. Fig. \ref{fig:alpha_poloidal} shows a composite picture of spatial variation of dynamo coefficients in the $(r,\theta)$ plane.
 Panels at the top, middle and bottom show the spatial distribution of diagonal, symmetric off-diagonal and 
 antisymmetric off-diagonal components of $\bf{a}$ tensor respectively. We put a black mask wherever the the error on the 
 coefficients calculated by the SVD method exceeds
 its absolute value. It is quite clear in Fig. \ref{fig:alpha_poloidal} that 
 SVD works quite well in extracting dynamo coefficients at high latitudes where large-scale dynamo dominates. At low latitudes, the fluctuation
 dynamo dominates the evolution, the mean fields and EMFs are small, and dynamo coefficients associated with the large-scale dynamo are vanishingly small.
 
 Regarding the distribution of dynamo coefficients in the poloidal plane, in a nutshell - apart from $\alpha_{\theta \theta}$, all other
 calculated dynamo coefficients show a coherent structure.  While $\alpha_{rr}$, $\alpha_{\phi \phi}$, $\alpha_{r \phi}$ and $\gamma_{\theta}$ show a  coherent antisymmetric behaviour about the mid-plane, $\alpha_{r \theta}$, $\alpha_{r \phi}$, $\alpha_{\theta \phi}$, $\gamma_{r}$ and $\gamma_{\phi}$ are coherently symmetric about the $\theta_l=0$ plane. Here, it must be mentioned that SVD treats each point in poloidal ($r,\theta$) plane independently. Therefore, the symmetry/antisymmetry of dynamo coefficients about the mid-plane is not an artefact of the SVD method.
 
 \subsection{Reliability check of the SVD method}
\label{sect:SVD_check}
The SVD method provides an estimate of error on the determination of the coefficients $a_{ij}$. It is evident from 
Fig. \ref{fig:alpha_poloidal} that at high latitudes, where mean fields, EMFs are stronger and dynamo
coefficients are of non-zero values, variances are small. However, where mean fields and EMFs are vanishingly small
(at low latitudes), errors are large (implying that a mean field description is invalid). In this subsection, we show some additional diagnostics to investigate the
accuracy of the parameterization as recovered by the SVD method.

\subsubsection{Reconstruction of the EMF}
\begin{figure*}
    %\centering
    \includegraphics[scale=0.45]{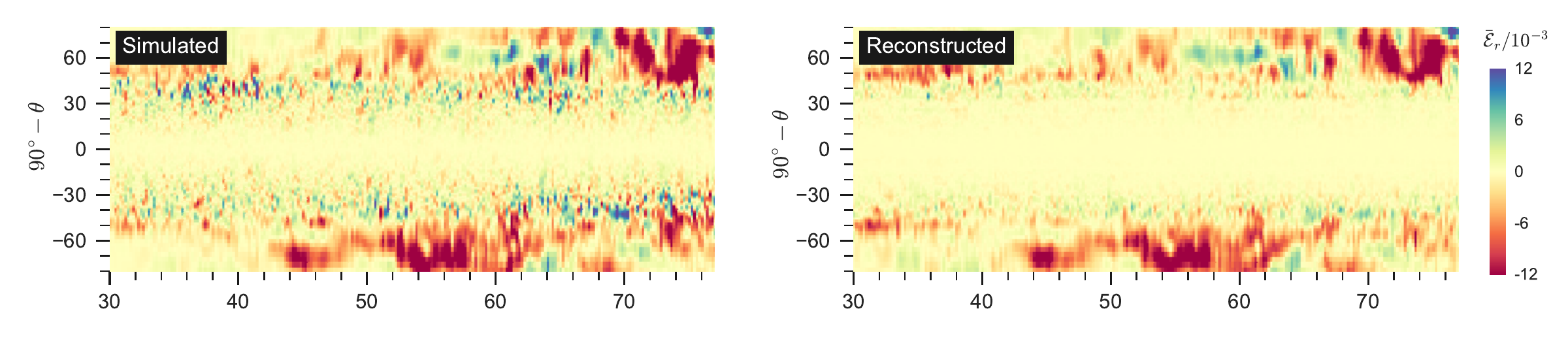}
    \includegraphics[scale=0.45]{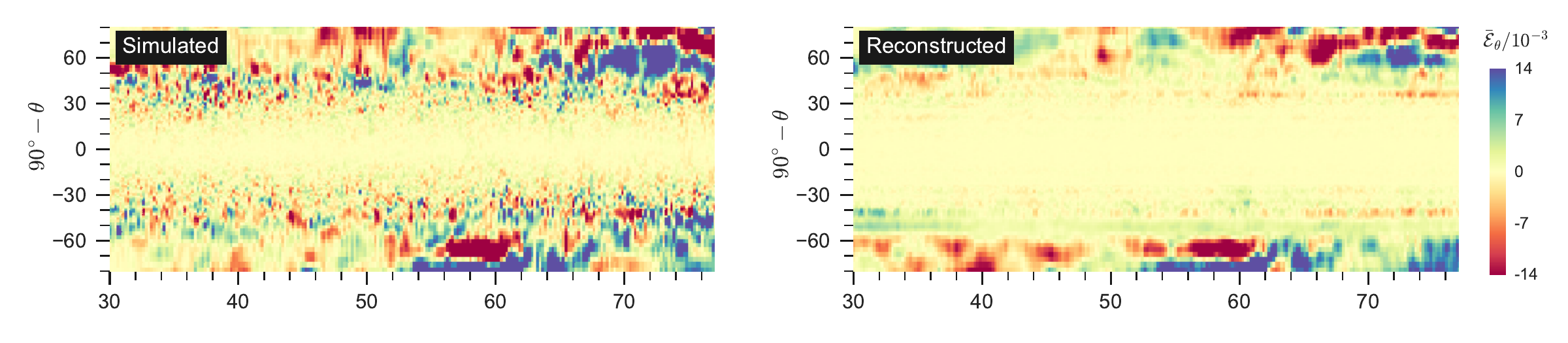}
    \includegraphics[scale=0.45]{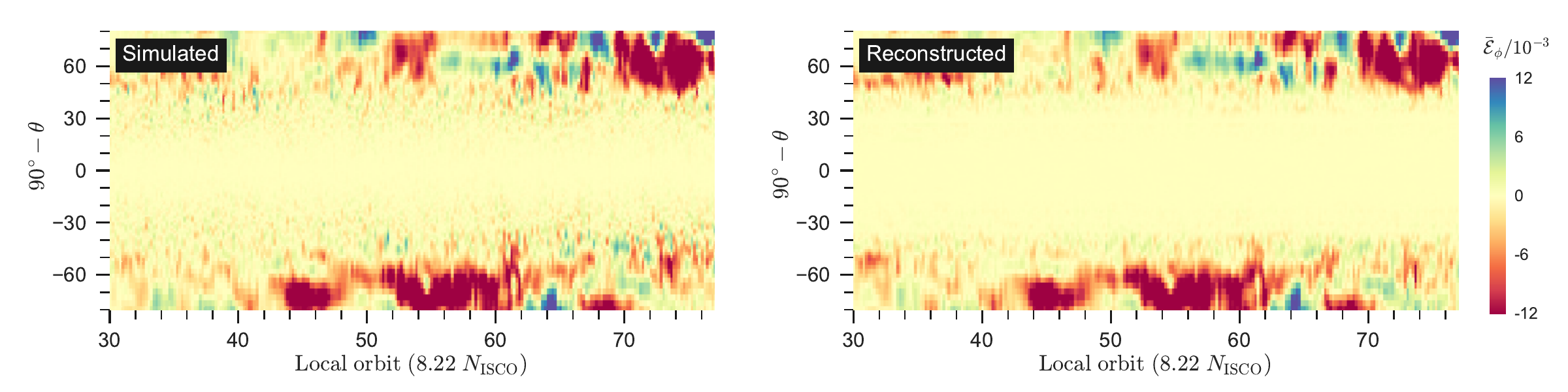}
 \caption{Spatio-temporal variation of radial (top panels), meridional (middle panels) and toroidal (bottom panels) components of  simulated and reconstructed EMFs at a radial distance $R_0=20$. 
Radial and toroidal components show a good agreement between the simulated and reconstructed EMFs at high latitudes, where 
dynamo coefficients are of non-zero values. At low latitudes, the match is poor because at low latitudes
large scale dynamo is suppressed and mean fields, EMFs are vanishingly small. The match between the simulated and reconstructed EMFs in the meridional component is not as good as for the other two components of EMFs.}
 \label{fig:emf_sim_recon}
 \end{figure*}

As a primary diagnostic, 
we reconstruct components of EMF using the extracted dynamo coefficients and mean magnetic fields (equation \ref{eq:emf_alpha_gamma}).
Fig. \ref{fig:emf_sim_recon} shows the comparison between the spatio-temporal variations of EMF (equation \ref{eq:mean_emf}) obtained directly from the simulation
(left panels) and reconstructed EMF $\bar{\mathcal{E}}^M$ (right panels) at a radius $r=20$. Top, middle and bottom panels of Fig. \ref{fig:emf_sim_recon} show the butterfly diagrams for radial ($\bar{\mathcal{E}}_r$), 
meridional ($\bar{\mathcal{E}}_{\theta}$) and toroidal ($\bar{\mathcal{E}}_{\phi}$) components of EMF respectively. 
It is evident that for $\bar{\mathcal{E}}_r$ and $\bar{\mathcal{E}}_{\phi}$ the match between the directly obtained EMF and 
reconstructed EMF is quite good at high latitudes, where the large-scale dynamo dominates. As expected, 
at low latitudes, the 
the reconstructed EMFs do not agree with the EMFs directly obtained from simulation because at low latitudes
large scale dynamo is suppressed and mean fields, EMFs are vanishingly small. For $\bar{\mathcal{E}}_{\theta}$, agreement between the
directly obtained and reconstructed EMFs is not as good as it is for $\bar{\mathcal{E}}_r$ and $\bar{\mathcal{E}}_{\phi}$. This poorer match  is again because of the smallness of $\bar{\mathcal{E}}_\theta$ and $\bar{B}_{\theta}$.

\subsubsection{Residuals of EMFs obtained from simulations and reconstructed one}
\begin{figure*}
    %\centering
    \includegraphics[scale=0.44]{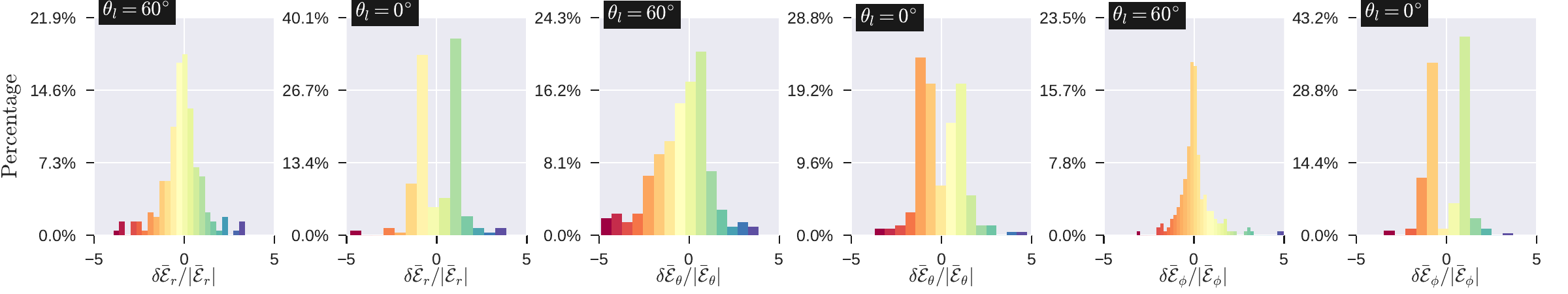}
 \caption{Histogram of the residuals of the simulated and reconstructed EMFs at two different latitudes $\theta_l=65^{\circ}$ (where large-scale dynamo dominates) and $\theta_l=0^{\circ}$ (where fluctuation dynamo operates)  at $r=20$.
Residual is divided by the magnitude of the simulated EMF. Note that the residuals are close to zero at high latitudes.}
   \label{fig:residual_hist}
 \end{figure*}
 
\begin{figure*}
    %\centering
    \includegraphics[scale=0.54]{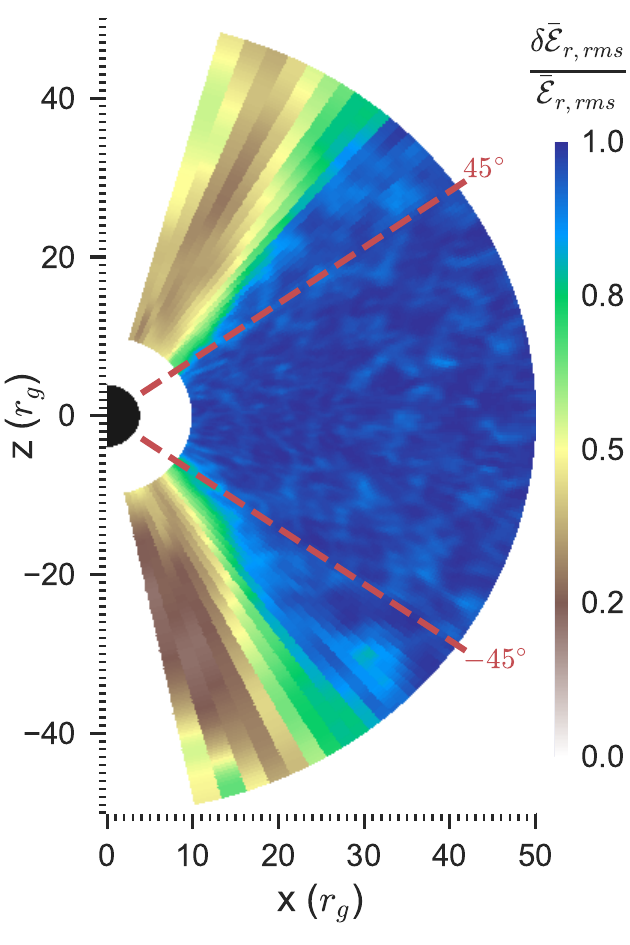}
    \includegraphics[scale=0.54]{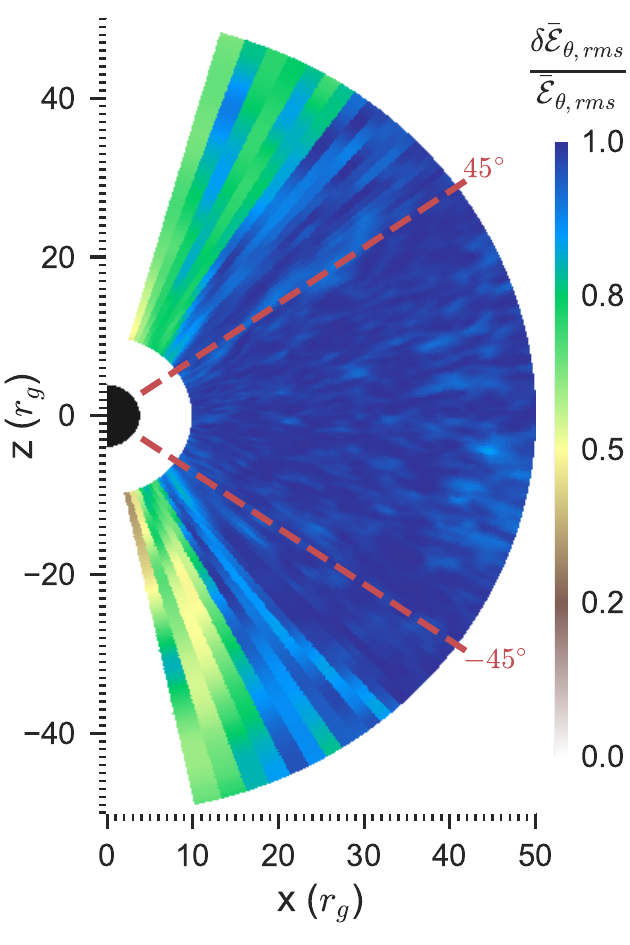}
    \includegraphics[scale=0.54]{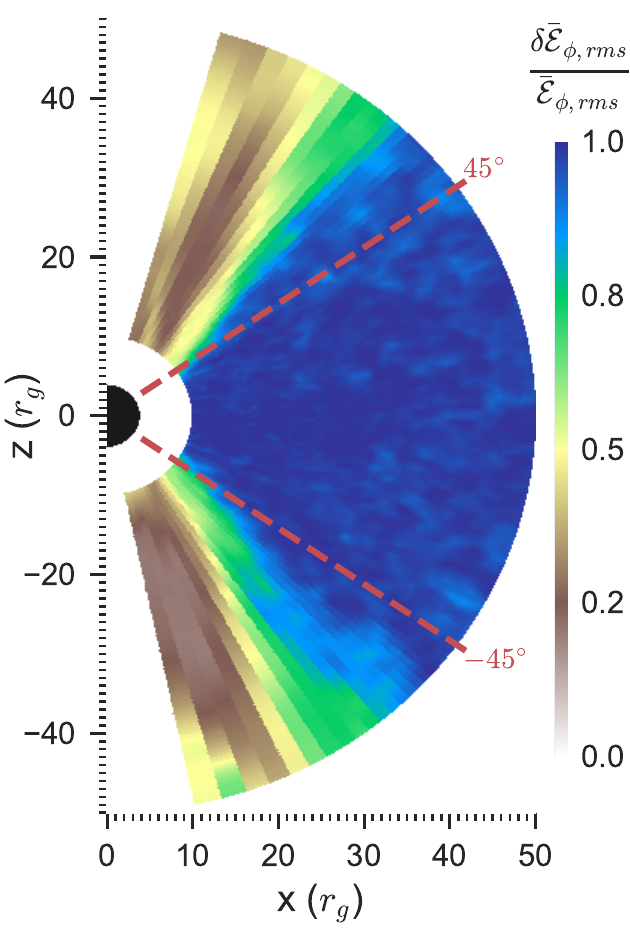}
 \caption{Distribution of rms values of the residuals between  the reconstructed and simulated EMFs in the poloidal ($r, \theta$) plane. Red dashed lines on northern and southern hemispheres denote the latitudes ($\theta_l=90^{\circ}-\theta$) $45^{\circ}$ and $-45^{\circ}$ respectively. The rms values of residuals are normalised by the rms values of the corresponding component of the EMF. Residuals are smaller for both radial and toroidal components beyond a scale-height, but comparatively larger for the meridional component.}
   \label{fig:emf_rms}
 \end{figure*}

Fig. \ref{fig:emf_sim_recon} gives a good visual impression of agreement between the EMFs obtained directly from simulations and reconstructed using the mean field closure.
To quantify the match, we calculate the residual of the directly obtained (from simulations) and reconstructed EMFs as follows
 \be
 \delta \bar{\mathcal{E}}_{i}(r,\theta,t)= \bar{\mathcal{E}}_{i}(r,\theta,t) - \bar{\mathcal{E}}^{M}_{i} (r,\theta,t). 
\ee
Fig. \ref{fig:residual_hist} shows the histogram of  residuals for each components of EMFs at a fixed radius $r=20$ and at two different latitudes- 
$\theta_l=60^{\circ}$ and $\theta_l=0^{\circ}$. The residuals are normalised by the respective components of EMFs $\bar{\mathcal{E}}_{i}(r,\theta,t)$.
It is easily observed that at high-latitudes, where the agreement is seen to be good in Fig. \ref{fig:emf_sim_recon}, the histogram shows a peak about zero.
At low latitudes (mid-plane), where  there is a large mis-match between  $\bar{\mathcal{E}}_{i}$ and  $\bar{\mathcal{E}}^{M}_{i}$ as seen in Fig. \ref{fig:emf_sim_recon}, there are few instances when the residuals are zero. 
It is also evident from the peak widths of the histograms that the match between $\bar{\mathcal{E}}_i$ and  
$\bar{\mathcal{E}}^{M}_{i}$  is best for $\phi$-component of EMF.

In Fig. \ref{fig:residual_hist}, we investigate the agreement between simulated (directly obtained from the simulation) and reconstructed EMFs at 
different times at particular point in space. 
To see the how well the match is in a time-averaged sense over the whole poloidal plane, we calculate at the rms values of the residual EMFs. 
Fig. \ref{fig:emf_rms} shows the poloidal distribution of residuals $\delta \bar{\mathcal{E}}_{i, rms}$. The residuals are normalised by the rms values of 
the EMFs $\bar{\mathcal{E}}_{i, rms}$. Fig. \ref{fig:emf_rms} essentially echoes the key results shown in Fig. \ref{fig:emf_sim_recon} and \ref{fig:residual_hist}.
We see that normalised residuals are minimum at high latitudes and maximum around the mid-plane. Thus SVD provides a good match between the reconstructed
and simulated EMFs at high latitudes where a large scale dynamo dominates. This also indicates that the $b_{ijk}$ tensor does not make
a statistically significant contribution to the EMF as we had stated earlier.
The match between the two EMFs
is poor at low latitudes as expected, where a turbulent small-scale dynamo operates.

\section{Discussion}
\label{sect:discussion}
\subsection{Comparison with previous local and global studies}
\label{sect:compare_works}
 All the previous global studies (\citealt{Arlt2001b, Flock2012a, Hogg2018a}) consider a simpler closure
 \be
 \label{eq:simple_alpha}
 \bar{\mathcal{E}}_{\phi} \approx \alpha_{\phi \phi} \bar{B}_{\phi}
 \ee
to characterise the dynamo coefficients in accretion flows. In this paper, we use a more general closure to calculate the diagonal as well as the off-diagonal dynamo coefficients in a global simulation of accretion flow around the black hole.  Determination of dynamo coefficients throughout the poloidal plane ($r, \theta$)  gives us the opportunity to compare our findings with the previous local and global studies.

We use the spherical polar coordinates ($r$, $\theta$, $\phi$) in our global simulations. Previous shearing box studies (\citealt{Brandenburg2008f, Gressel2010, Gressel2015}) characterising dynamo coefficients use the TF method. For  comparison, we can make a mapping from spherical polar to Cartesian coordinates for the three dynamo coefficients; $\alpha_{xx} \to \alpha_{rr}$,
$\alpha_{yy} \to \alpha_{\phi\phi}$ and $\gamma_{z} \to \gamma_{\theta}$. The trend in the meridional variation of these  three coefficients agrees well between local 
shearing box simulations (see Fig. 5 of \cite{Gressel2010}) and our global study (see Fig. \ref{fig:alpha_vertical}). For example, like $\alpha_{yy}$, its spherical counterpart $\alpha_{\phi \phi}$
is  negative / positive close the mid-plane and tends to be positive/ negative at higher latitudes in the northern hemisphere (NH)/southern hemisphere (SH).  However, to be confident about
prevalence of  positive sign at high latitudes, the calculation needs to be done for a geometrically thin disc ($H/R \ll 1$), where many scale-heights are available.

There is a long standing disagreement between local and global simulations regarding the sign of $\alpha_{\phi \phi}$, calculated using the simple closure (equation \ref{eq:simple_alpha}). Local studies (\citealt{Brandenburg1995, Brandenburg1997, Davis2010}) found a negative $\alpha_{\phi \phi}$ in NH. Global studies (\citealt{Arlt2001b, Flock2012a}) found an opposite trend, i.e a  positive $\alpha_{\phi \phi}$ in  NH.  However, \cite{Hogg2018a} obtained an $\alpha_{\phi \phi}$ of  negative sign at $r=15$  
and between $+H$ to $+2H$ (i.e in NH) using the same simple closure (equation \ref{eq:simple_alpha}).
The reason behind the agreement/disagreement among the different studies especially among the recent global studies (\citealt{Flock2012a, Hogg2018a} and the current work) is not very clear. All  three works use the same code ({\tt PLUTO}) and similar combination of algorithms recommended  by \cite{Flock2010}.

\subsection{Opportunity to build an effective mean field model}
To attain saturation of the MRI  and for dynamo action, an expensive 3D simulation has to be performed 
for a sufficiently long time. Determination of the spatial profiles of dynamo coefficients 
gives an opportunity to build an effective mean field model for 
MRI driven accretion flow, which can be run in 2D with sustained turbulence. In particular, an effective mean field 
model can be useful in simulating the expensive geometrically thin disc ($H/R \ll 1$) with a long dynamical range.

There have been previous attempts to build an effective mean field models for MRI driven accretion flow (\citealt{Bucciantini2013, Stepanovs2014, Scadowski2015, Fendt2018, Tomei2020}). However, all those models  consider  either an $\alpha$-effect or both $\alpha$-effect and turbulent diffusivity using some simple closure. 
Consideration of the full set of dynamo coefficients in the mean field model is expected to provide a more realistic result.

\subsection{Characterisation of the dynamo}
\label{sect:which_dynamo}
\begin{figure}
    %\centering
    \includegraphics[scale=0.55]{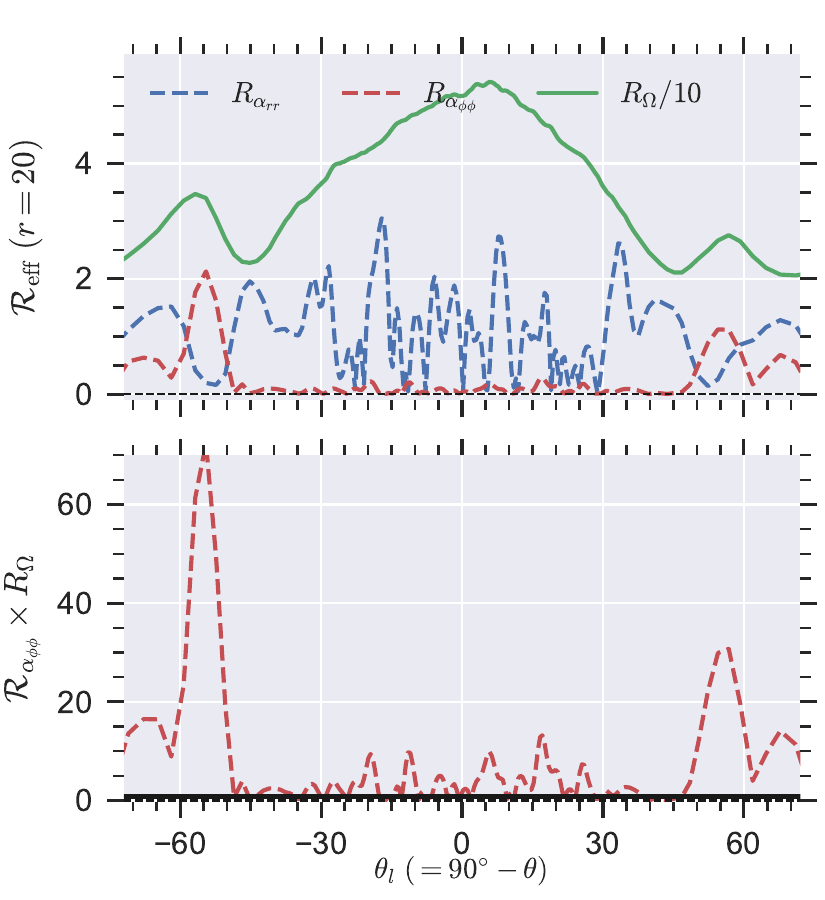}
 \caption{Top panel: Meridional variation of the control parameters related to the measurement of the strength of $\alpha$-effect (equations \ref{eq:R_alpha}) 
and  shear (\ref{eq:R_omega}) at $r=20$.  Bottom panel: Meridional variation of the dynamo number $D=R_{\alpha_{\phi \phi}} R_{\Omega}$ at $r=20$.}
   \label{fig:reynolds_no}
 \end{figure}
Dynamo action in the differentially rotating  accretion disc is traditionally categorised  as an $\alpha-\Omega$ dynamo (\citealt{Brandenburg1995, Gressel2010}). Here, $\Omega$ effect refers to the generation of toroidal component of magnetic field
from the poloidal components by differential rotation. Poloidal field can be regenerated from the toroidal field by an $\alpha$-effect. 
The coefficient $\alpha_{\phi \phi}$ is responsible for the conversion of toroidal fields into poloidal fields. The coefficients $\alpha_{rr}$ and $\alpha_{\theta \theta}$ are related to generation of toroidal fields by the $\alpha$-effect. However, the regeneration of the toroidal field from poloidal field is dominated
by strong shear, or the $\Omega$-effect in accretion discs. Therefore these dynamos are referred to as $\alpha$-$\Omega$ dynamos.

 We quantify the strength of such a dynamo by using the mean-field induction equation to identify different dimensionless control parameters. Firstly, the parameter $R_{\alpha_{ij}} = \alpha_{ij} H/\eta_T$ measures
the importance of the $\alpha$-effect to amplify the field against turbulent diffusion $\eta_T$. The importance of
the shear is governed by the parameter $R_{\Omega}=SH^2/\eta_T$. The product $D=R_{\alpha_{\phi \phi}} R_\Omega$ called the dynamo
number, measures the combined amplification strength of both
the $\alpha$ and $\Omega$ effects in an $\alpha$-$\Omega$ dynamo. 
It should exceed a critical value $D_{crit}$ for amplification
of the field, which for thin discs is $\sim 10$ 
\citep{Brandenburg2005}. Of course the value of $D_{crit}$ for the RIAF is not known apriori. The value of shear $S=1.5 \Omega$ and scale-height $H=c_s/\Omega$, are known
from the simulation data, while $\alpha_{ij}$ can be obtained
from the SVD analysis. 
Turbulent diffusion is approximated as $\eta_T =  v^{\prime}_{rms}/3k_{\rm MRI}$, where we are motivated from the TF result on forced turbulence \citep{Sur2008}, with $k_{MRI}=\Omega/v_A$, the wavenumber of the 
fastest growing mode of MRI. This leads to an estimate,
 \be
 \label{eq:R_alpha}
R_{\alpha_{ij}} = \frac{\alpha_{ij} H}{\eta_T}=3\sqrt{\frac{\beta}{2}} \frac{\alpha_{ij}}{v^{\prime}_{rms}}, 
\ee

 \be
 R_{\Omega}=\frac{SH^2}{\eta_T}=\frac{9}{4}\beta \frac{v_A}{v^{\prime}_{rms}},
 \label{eq:R_omega}
 \ee
 where $\beta=2P_{\rm gas}/B^2$. 

 These numbers are generally studied in the kinematic stage; but here we calculate them here in saturation.
Top and bottom panels of Fig. \ref{fig:reynolds_no} show the meridional variation of the control parameters ($R_{\alpha_{\phi \phi}}$, $R_{\alpha_{rr}}$ and $R_{\Omega}$)
and dynamo number $D= R_{\alpha_{\phi \phi}}R_{\Omega}$ with latitude respectively. 
Here, we consider only $\alpha_{rr}$ and $\alpha_{\phi \phi}$, related to the dominant components of magnetic field $B_r$ and $B_{\phi}$ respectively.
 It is evident from Fig. \ref{fig:reynolds_no}  that at all latitudes $R_{\Omega}>R_{\alpha_{ij}}$ so that dynamo generation is controlled by the dynamo number
$D=R_{\alpha_{\phi \phi}}R_{\Omega}$. 
This is of course a local value of $D$ at some $(r,\theta)$ and 
to infer whether the dynamo is super critical, one needs to solve the mean-field induction equation with these $\alpha_{ij}$ and shear profiles. But it is interesting to note from the the meridional profile of the local value of $D$, that it exceeds about 10 where the mean field is prominent. More work solving the mean-field dynamo equation is needed to put this conclusion on a firmer footing. We expect the dynamo action to be local because both the kinetic and current helicities become non-vanishing at high latitudes (Fig. 26 in \cite{Dhang2019}) where a large-scale dynamo operates and dynamo coefficients have non-zero values.

\subsection{Turbulent pumping, radially outward transport of the mean field}
\label{sect:discuss_pumping}

\begin{figure}
    %\centering
    \includegraphics[scale=0.65]{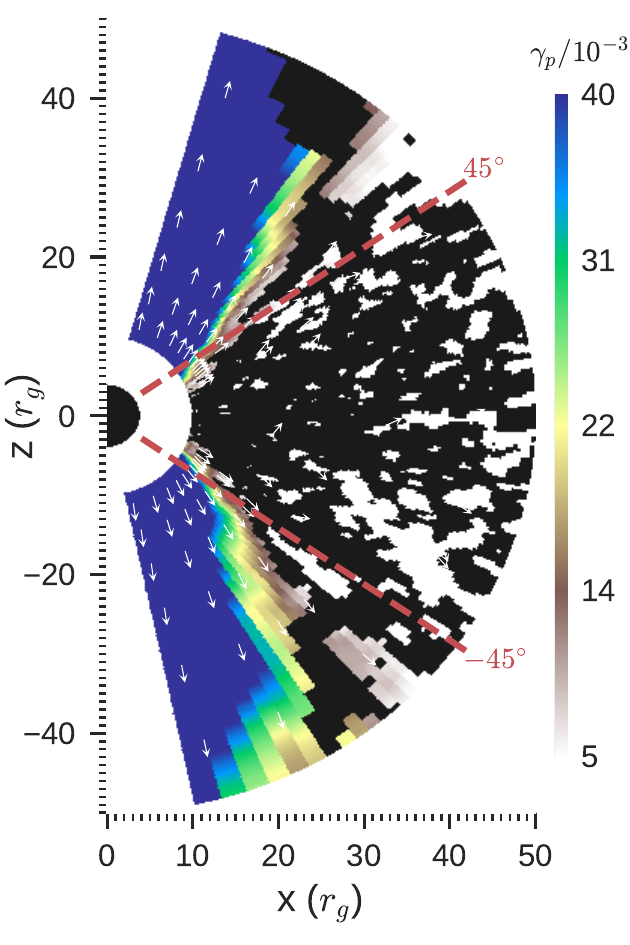}
 \caption{Distribution of $\boldsymbol{\gamma_p}  = \gamma_r \hat{r} + \gamma_{\theta} \hat{\theta}$ in the poloidal plane. Colour and arrows denote the magnitude and direction of $\gamma_p$ respectively. A black mask is applied in the regions where the ratio of ratio of standard deviation to magnitude of the quantity is greater than the unity.    }
   \label{fig:gamma_quiver}
 \end{figure} 
 
\begin{figure}
    %\centering
    \includegraphics[scale=0.55]{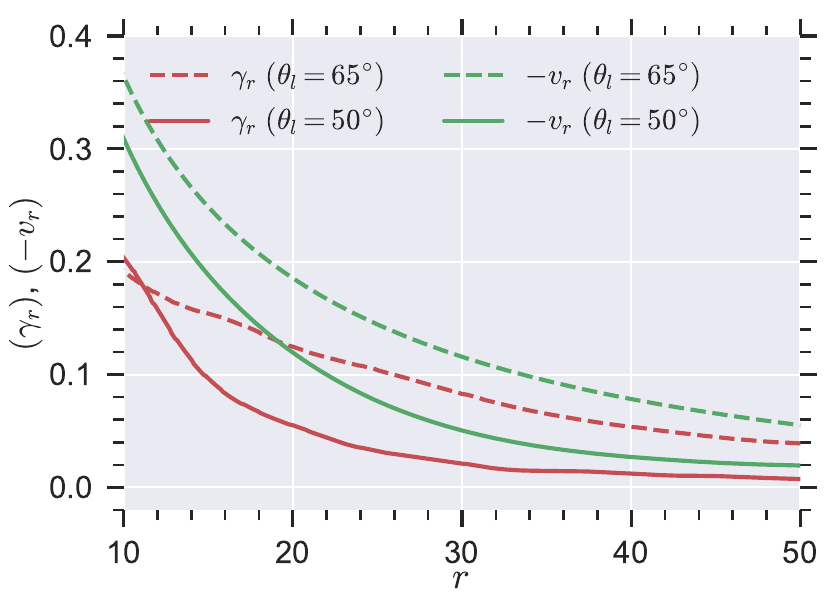}
 \caption{Comparison of radial turbulent pumping $\gamma_r  (r)$ and radial advection $\bar{v}_{r}(r)$ at two different latitudes $\theta_l=50^{\circ}, 65^{\circ}$. 
Radially outward turbulent pumping is significant enough to oppose the inward advection of magnetic flux by the mean flow.     }
   \label{fig:gamma_vs_vr}
 \end{figure}
 Studies (\citealt{Lubow1994, Guilet2013}) which consider the accretion of a large-scale poloidal field on to the central
black hole consider turbulent diffusion as the only  factor opposing  
inward advection of magnetic flux. Importance of the turbulent pumping in mean magnetic flux transport was previously unnoticed. In our study, we find that along with turbulent diffusion, turbulent pumping can also  act against  inward advection. Fig.\ref{fig:alpha_poloidal} shows $\gamma_{\theta}$ that represents the transport 
of the mean magnetic field from turbulent region close to the mid-plane to the less turbulent coronal region. On the other hand, $\gamma_r$ shows that
the  mean field is transported radially outward. The direction of transport is down the gradient of turbulence intensity which we find decreases with 
increasing radius, as $v^{\prime}_{rms} \sim r^{-1}$.

 To see the net effect of these two coefficients, we construct a vector quantity 
$\boldsymbol{\gamma_p}  = \gamma_r \hat{r} + \gamma_{\theta} \hat{\theta}$. Fig \ref{fig:gamma_quiver} shows the distribution of $\boldsymbol{\gamma_p}$
in the poloidal plane. Colour describes the magnitude and arrow, the direction. A black mask is applied wherever the standard deviation is larger than the magnitude
of $\boldsymbol{\gamma_p}$ . Fig. \ref{fig:gamma_quiver} clearly shows that the large scale field is transported radially outward due to turbulent diamagnetism at high latitudes. Fig. \ref{fig:gamma_vs_vr} shows the comparison of radial profiles of turbulent pumping $\gamma_r$
and radial advection $\bar{v}_r$ at two different latitudes $\theta_l=55^{\circ}, 65^{\circ}$. At all radii, $\gamma_r$ is comparable to $\bar{v}_r$, the difference decreases with an increasing radius. We expect the effect of turbulent pumping to be  stronger in case of a geometrically thin disc ($H/R \ll 1$) where the radial velocity (even in the coronal region) is smaller.
  
Recent net flux simulations  of a geometrically thin Keplerian disc also show that magnetic flux does not accumulate efficiently near the centre due 
to the quasi-steady balance of advection and diffusion of magnetic fields (\citealt{Zhu2018, Mishra2019a}). 
Our study suggests that this may be due to the presence of a strong turbulent pumping rather than simply turbulent diffusion. However, flux accumulation can be efficient if the RIAF is magnetically arrested (MAD; \citealt{Bisnovatyi-Kogan1974a, Igumenshchev2003, Narayan2003, Tchekhovskoy2011, McKinney2012}). 
In case of a MAD, the strong coherent large-scale magnetic field is dynamically important and the dynamo is completely suppressed (e.g. see  Fig. 16 of \cite{McKinney2012}) 
and the large scale field is efficiently dragged in.

\subsection{Dynamo action and the jet connection}

\begin{figure*}
    %\centering
    \includegraphics[scale=0.5]{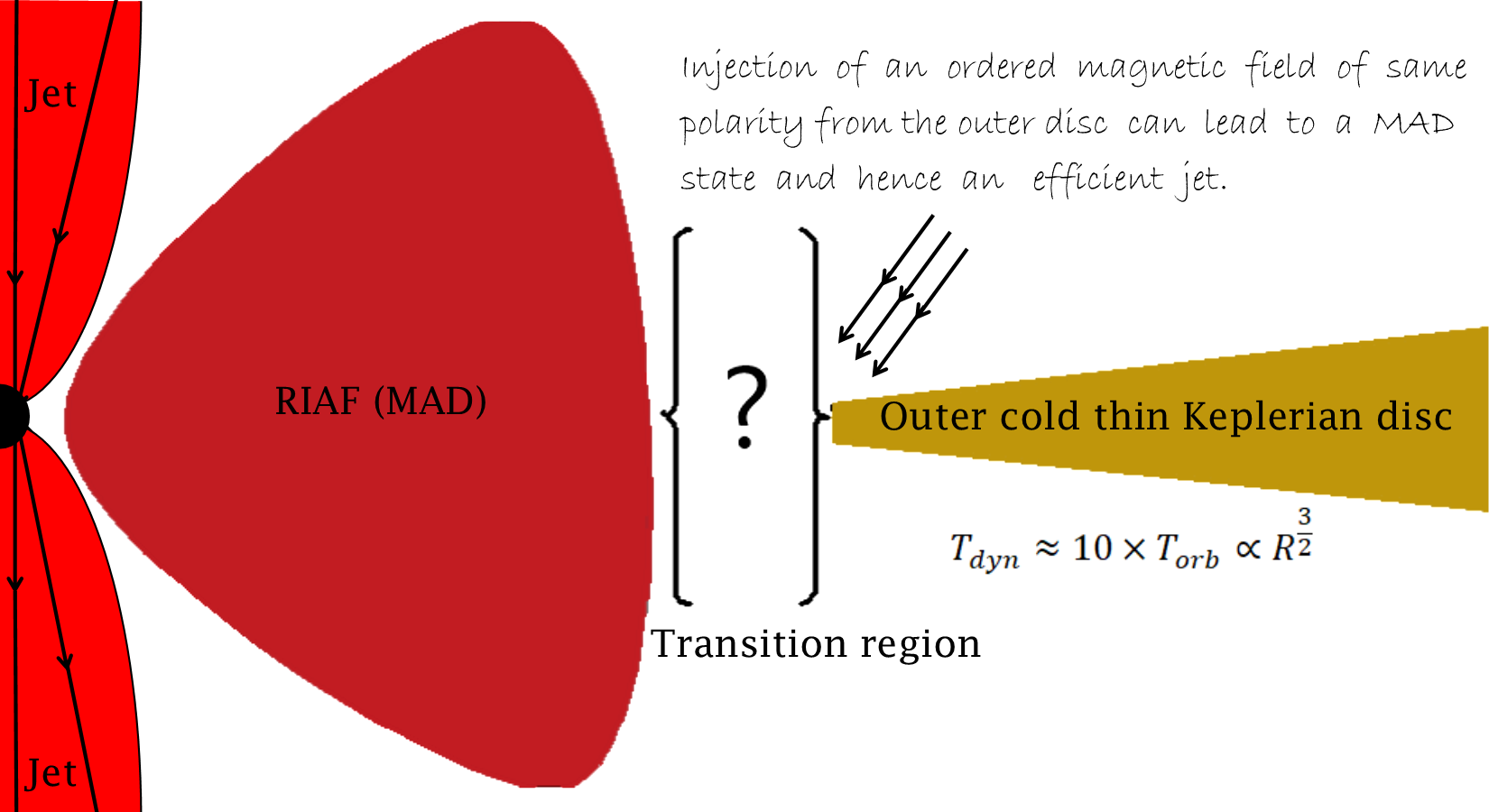}
 \caption{Cartoon diagram of the geometry of the accretion flow in the low/hard state of BHBs in the truncated disc model. The model requires 
a outer  cold, optically thick, geometrically thin disc and an inner hot, optically thin, geometrically thick flow (RIAF). While jets are present in the 
low/hard state, they are absent in the high/soft state. Dynamo generated large-scale ($\sim H \propto r$) field produced at the inner edge
of the outer thin disc can be supplied to the inner RIAF to establish a magnetically dominated RIAF that aids formation of jets.   }
   \label{fig:cartoon_jet}
\end{figure*}
Jets are observed in hard spectral state in BHBs. 
The basic requirements to produce the jet are a sufficiently strong, ordered poloidal magnetic field and an inner hot geometrically thick accretion flow (e.g. see \cite{Meier2005c}).
However, numerical simulations show that there is no apparent correlation between the disc scale height and jet power (\citealt{Fragile2012a}). This leads to the possibility that magnetic field geometry might be the deciding factor in determining the jet power, 
although, association of jets with low/hard state implies the presence of a RIAF close to the black hole (\citealt{Fender1999a}).
In section \ref{sect:discuss_pumping}, we discussed the difficulty
in transporting large-scale magnetic fields inward close to the black hole unless accretion occurs in the magnetically dominated regime. 

In this section, we propose a scenario, in which a large scale mean field dynamo operates in the outer thin disc and seeds a MAD state in the inner RIAF.\footnote{Although turbulent pumping gives outward transport of mean fields, the mean inflow speed dominates and causes inward advection of a large-scale poloidal field in the coronal region (Figure \ref{fig:gamma_vs_vr}).} This magnetic arrangement is conducive to jet launching.
Fig. \ref{fig:cartoon_jet} shows the accretion flow geometry in the truncated disc model (\citealt{Esin1997, Done2007}). According to this model, 
the outer standard disc is truncated to join with an inner RIAF over a transition radius $r_{tr}\approx 30-200$ $r_g$ (\citealt{Nemmen2014}). 
In the outer thin disc, dynamo action is strong, and magnetic fields flip with a dynamo period $T_{\rm dyn}\approx 10-15 T_{\rm orb} \propto r^{3/2}$ which is  much longer than the timescale at ISCO where a jet is launched. Hence, {\em innermost part} of the outer standard disc can supply coherent large-scale ($\sim H$) dynamo generated ordered magnetic fields to the inner RIAF leading to a  magnetically dominated accretion flow and hence a favourable condition for jets.

It should be mentioned that the generation of a MAD depends on the size of the inner RIAF. If area occupied by the inner RIAF shrinks
and inner disc moves inward, the polarity of the supplied field flips more quickly as $T_{\rm dyn} \propto r^{3/2}$. Therefore, time span over which 
mean fields of the same polarity is supplied to inner RIAF is shorter. The supply of opposite polarity fields enhances reconnection within the RIAF and the 
disc becomes less magnetised and can lead to the transient 
ballistic jets from the region close to the black hole (\citealt{Dexter2014a}). However, few studies (e.g \citealt{Stepanovs2014}) associate episodic jets with the duty cycles of the dynamo in the thin disc itself. In the soft state, a thin disc extends till the ISCO. Efficient dynamo action in the thin disc close to the black hole leads to the frequent polarity reversal of the dynamo generated large scale magnetic fields. Hence, the jet is quenched due to the weakening of coherent large scale poloidal fields.

\section{Summary}
\label{sect:summary}

In this paper, we have analysed 
the MRI driven dynamo in a weakly magnetised radiatively inefficient accretion flow (RIAF). We use the  mean field dynamo formalism to understand mechanism for the generation of large-scale magnetic fields in the RIAF (model M-2P discussed in \cite{Dhang2019}).
The key results of our work are summarised below.
\begin{itemize}
\item We have recovered the dynamo coefficients for the MRI driven large-scale dynamo in the RIAF using the
SVD method. Our study is the first one to 
calculate the distribution of dynamo coefficients in the poloidal plane ($r, \theta$). Out of the calculated coefficients, four coefficients
$\alpha_{rr}$, $\alpha_{\phi \phi}$, $\alpha_{r \phi}$ and $\gamma_{\theta}$ are anti-symmetric about the mid-plane 
($\theta_l=90^{\circ}-\theta=0^{\circ}$). Rest of the coefficients ($\alpha_{r \theta}$, $\alpha_{\theta \phi}$, $\gamma_r$ and $\gamma_{\phi}$, $\alpha_{\theta \theta}$)
 show symmetric behaviour about the mid-plane. Many of the the calculated coefficients roughly scale as a  power law $\propto r^{-1.5}$, similar to the angular velocity $\Omega \propto r^{-1.7}$.

\item We find that meridional variations of $\alpha_{\phi \phi}$, responsible for toroidal to poloidal field conversion and considered to be the most important dynamo coefficient,
 is very similar to that found in shearing box simulations using the `test field method'.
 %and similar to our previous simplified estimate (\citealt{Dhang2019})}.
 The dynamo coefficient $\alpha_{\phi \phi}$ is  negative at low latitudes and tends to be  positive at higher latitudes.

\item We estimate the relative importance of the $\alpha$-effect and shear. We conclude that although $\alpha_{rr}$ is quite large, 
the MRI driven large-scale dynamo is essentially
of the $\alpha-\Omega$ type.

\item The information of dynamo coefficients will allow us to carry out less expensive axisymmetric  global accretion 
disc simulations with sustained turbulence. Simulating such an 
effective mean field model is especially attractive in context of geometrically thin disc with a large dynamical range.

\item We find  a strong turbulent pumping, which transports large scale magnetic fields 
radially outward. This effect along with the turbulent diffusion
makes  it difficult for the large-scale magnetic fields, an important factor to produce jets,  to be advected inward by the mean flow.

\item We propose a mechanism to explain the presence of jets in the low/hard state of a BHB considering a truncated disc model.
We outline a scenario where a dynamo generated large-scale 
field produced at the inner edge
of the outer thin disc can be supplied to the inner RIAF to establish a magnetically dominated RIAF (MAD) conducive to jet formation.

\end{itemize}

\section*{Acknowledgements}
PD would like to thank Xue-Ning Bai, Paul Charbonneau and Bhupendra Mishra for useful discussions. We thank Oliver Gressel for detailed comments on the manuscript. We thank the reviewer for the useful comments. PD would also like to thank IUCAA for the hospitality extended to him during his multiple visits to the institute. PS acknowledges a Swarnajayanti Fellowship from the Department of Science and Technology (DST/SJF/PSA- 03/2016-17). The simulations analysed in this paper were carried out on Cray XC40-SahasraT cluster at Supercomputing Education and Research Centre (SERC), IISc.

\bibliographystyle{mnras}
\bibliography{bibtex_dynamo}

\begin{thebibliography}{}
\makeatletter
\relax
\def\mn@urlcharsother{\let\do\@makeother \do\$\do\&\do\#\do\^\do\_\do\%\do\~}
\def\mn@doi{\begingroup\mn@urlcharsother \@ifnextchar [ {\mn@doi@}
  {\mn@doi@[]}}
\def\mn@doi@[#1]#2{\def\@tempa{#1}\ifx\@tempa\@empty \href
  {http://dx.doi.org/#2} {doi:#2}\else \href {http://dx.doi.org/#2} {#1}\fi
  \endgroup}
\def\mn@eprint#1#2{\mn@eprint@#1:#2::\@nil}
\def\mn@eprint@arXiv#1{\href {http://arxiv.org/abs/#1} {{\tt arXiv:#1}}}
\def\mn@eprint@dblp#1{\href {http://dblp.uni-trier.de/rec/bibtex/#1.xml}
  {dblp:#1}}
\def\mn@eprint@#1:#2:#3:#4\@nil{\def\@tempa {#1}\def\@tempb {#2}\def\@tempc
  {#3}\ifx \@tempc \@empty \let \@tempc \@tempb \let \@tempb \@tempa \fi \ifx
  \@tempb \@empty \def\@tempb {arXiv}\fi \@ifundefined
  {mn@eprint@\@tempb}{\@tempb:\@tempc}{\expandafter \expandafter \csname
  mn@eprint@\@tempb\endcsname \expandafter{\@tempc}}}

\bibitem[\protect\citeauthoryear{Arlt \& R{\"u}diger}{Arlt \&
  R{\"u}diger}{2001}]{Arlt2001b}
Arlt R.,  R{\"u}diger G.,  2001, \mn@doi [\aap] {10.1051/0004-6361:20010797},
  374, 1035

\bibitem[\protect\citeauthoryear{Bai \& Stone}{Bai \& Stone}{2013}]{Bai2013}
Bai X.-N.,  Stone J.~M.,  2013, \mn@doi [\apj] {10.1088/0004-637X/769/1/76},
  769, 76

\bibitem[\protect\citeauthoryear{Balbus \& Hawley}{Balbus \&
  Hawley}{1991}]{Balbus1991}
Balbus S.~A.,  Hawley J.~F.,  1991, \mn@doi [\apj] {10.1086/170270}, 376, 214

\bibitem[\protect\citeauthoryear{Bendre, Subramanian, Elstner  \&
  Gressel}{Bendre et~al.}{2019}]{Bendre2019}
Bendre A.~B.,  Subramanian K.,  Elstner D.,   Gressel O.,  2019, \mn@doi
  [\mnras] {10.1093/mnras/stz3267}

\bibitem[\protect\citeauthoryear{{Bhat}, {Ebrahimi}  \& {Blackman}}{{Bhat}
  et~al.}{2016}]{Bhat2016}
{Bhat} P.,  {Ebrahimi} F.,   {Blackman} E.~G.,  2016, \mn@doi [\mnras]
  {10.1093/mnras/stw1619}, \href
  {https://ui.adsabs.harvard.edu/abs/2016MNRAS.462..818B} {462, 818}

\bibitem[\protect\citeauthoryear{Bisnovatyi-Kogan \&
  Ruzmaikin}{Bisnovatyi-Kogan \& Ruzmaikin}{1974}]{Bisnovatyi-Kogan1974a}
Bisnovatyi-Kogan G.~S.,  Ruzmaikin A.~A.,  1974, \mn@doi [\apss]
  {10.1007/BF00642237}, 28, 45

\bibitem[\protect\citeauthoryear{Bisnovatyi-Kogan \&
  Ruzmaikin}{Bisnovatyi-Kogan \& Ruzmaikin}{1976}]{Bisnovatyi-Kogan1976}
Bisnovatyi-Kogan G.~S.,  Ruzmaikin A.~A.,  1976, \mn@doi [\apss]
  {10.1007/BF01225967}, 42, 401

\bibitem[\protect\citeauthoryear{{Blackman} \& {Nauman}}{{Blackman} \&
  {Nauman}}{2015}]{Blackman2015}
{Blackman} E.~G.,  {Nauman} F.,  2015, \mn@doi [Journal of Plasma Physics]
  {10.1017/S0022377815000999}, \href
  {https://ui.adsabs.harvard.edu/abs/2015JPlPh..81e3905B} {81, 395810505}

\bibitem[\protect\citeauthoryear{Blandford \& Begelman}{Blandford \&
  Begelman}{1999}]{Blandford1999}
Blandford R.~D.,  Begelman M.~C.,  1999, \mn@doi [\mnras]
  {10.1046/j.1365-8711.1999.02358.x}, 303, L1

\bibitem[\protect\citeauthoryear{Blandford \& Payne}{Blandford \&
  Payne}{1982}]{Blandford1982}
Blandford R.~D.,  Payne D.~G.,  1982, \mn@doi [\mnras]
  {10.1093/mnras/199.4.883}, 199, 883

\bibitem[\protect\citeauthoryear{Blandford \& Znajek}{Blandford \&
  Znajek}{1977}]{Blandford1977}
Blandford R.~D.,  Znajek R.~L.,  1977, \mn@doi [\mnras]
  {10.1093/mnras/179.3.433}, 179, 433

\bibitem[\protect\citeauthoryear{Brandenburg}{Brandenburg}{2008}]{Brandenburg2008f}
Brandenburg A.,  2008, \mn@doi [Astronomische Nachrichten]
  {10.1002/asna.200811027}, 329, 725

\bibitem[\protect\citeauthoryear{Brandenburg}{Brandenburg}{2009}]{Brandenburg2009d}
Brandenburg A.,  2009, \mn@doi [\ssr] {10.1007/s11214-009-9490-0}, 144, 87

\bibitem[\protect\citeauthoryear{Brandenburg \& Donner}{Brandenburg \&
  Donner}{1997}]{Brandenburg1997}
Brandenburg A.,  Donner K.~J.,  1997, \mn@doi [\mnras]
  {10.1093/mnras/288.2.L29}, 288, L29

\bibitem[\protect\citeauthoryear{Brandenburg \& Subramanian}{Brandenburg \&
  Subramanian}{2005}]{Brandenburg2005}
Brandenburg A.,  Subramanian K.,  2005, \mn@doi [\physrep]
  {10.1016/j.physrep.2005.06.005}, 417, 1

\bibitem[\protect\citeauthoryear{Brandenburg, Nordlund, Stein  \&
  Torkelsson}{Brandenburg et~al.}{1995}]{Brandenburg1995}
Brandenburg A.,  Nordlund A.,  Stein R.~F.,   Torkelsson U.,  1995, \mn@doi
  [\apj] {10.1086/175831}, 446, 741

\bibitem[\protect\citeauthoryear{Bucciantini \& Del~Zanna}{Bucciantini \&
  Del~Zanna}{2013}]{Bucciantini2013}
Bucciantini N.,  Del~Zanna L.,  2013, \mn@doi [\mnras] {10.1093/mnras/sts005},
  428, 71

\bibitem[\protect\citeauthoryear{{Cao}}{{Cao}}{2018}]{Cao2018}
{Cao} X.,  2018, \mn@doi [\mnras] {10.1093/mnras/stx2688}, \href
  {https://ui.adsabs.harvard.edu/abs/2018MNRAS.473.4268C} {473, 4268}

\bibitem[\protect\citeauthoryear{Chandrasekhar}{Chandrasekhar}{1960}]{Chandrasekhar1960a}
Chandrasekhar S.,  1960, \mn@doi [Proceedings of the National Academy of
  Science] {10.1073/pnas.46.2.253}, 46, 253

\bibitem[\protect\citeauthoryear{Davis, Stone  \& Pessah}{Davis
  et~al.}{2010}]{Davis2010}
Davis S.~W.,  Stone J.~M.,   Pessah M.~E.,  2010, \mn@doi [\apj]
  {10.1088/0004-637X/713/1/52}, 713, 52

\bibitem[\protect\citeauthoryear{Dexter, McKinney, Markoff  \&
  Tchekhovskoy}{Dexter et~al.}{2014}]{Dexter2014a}
Dexter J.,  McKinney J.~C.,  Markoff S.,   Tchekhovskoy A.,  2014, \mn@doi
  [\mnras] {10.1093/mnras/stu581}, 440, 2185

\bibitem[\protect\citeauthoryear{Dhang \& Sharma}{Dhang \&
  Sharma}{2019}]{Dhang2019}
Dhang P.,  Sharma P.,  2019, \mn@doi [\mnras] {10.1093/mnras/sty2692}, 482, 848

\bibitem[\protect\citeauthoryear{Done, Gierli{\'n}ski  \& Kubota}{Done
  et~al.}{2007}]{Done2007}
Done C.,  Gierli{\'n}ski M.,   Kubota A.,  2007, \mn@doi [\aapr]
  {10.1007/s00159-007-0006-1}, 15, 1

\bibitem[\protect\citeauthoryear{Esin, McClintock  \& Narayan}{Esin
  et~al.}{1997}]{Esin1997}
Esin A.~A.,  McClintock J.~E.,   Narayan R.,  1997, \mn@doi [\apj]
  {10.1086/304829}, 489, 865

\bibitem[\protect\citeauthoryear{Fender et~al.,}{Fender
  et~al.}{1999}]{Fender1999a}
Fender R.,  et~al., 1999, \mn@doi [\apjl] {10.1086/312128}, 519, L165

\bibitem[\protect\citeauthoryear{{Fendt} \& {Ga{\ss}mann}}{{Fendt} \&
  {Ga{\ss}mann}}{2018}]{Fendt2018}
{Fendt} C.,  {Ga{\ss}mann} D.,  2018, \mn@doi [\apj]
  {10.3847/1538-4357/aab14c}, \href
  {https://ui.adsabs.harvard.edu/abs/2018ApJ...855..130F} {855, 130}

\bibitem[\protect\citeauthoryear{Flock, Dzyurkevich, Klahr  \& Mignone}{Flock
  et~al.}{2010}]{Flock2010}
Flock M.,  Dzyurkevich N.,  Klahr H.,   Mignone A.,  2010, \mn@doi [\aap]
  {10.1051/0004-6361/200912443}, 516, A26

\bibitem[\protect\citeauthoryear{Flock, Dzyurkevich, Klahr, Turner  \&
  Henning}{Flock et~al.}{2012}]{Flock2012a}
Flock M.,  Dzyurkevich N.,  Klahr H.,  Turner N.,   Henning T.,  2012, \mn@doi
  [\apj] {10.1088/0004-637X/744/2/144}, 744, 144

\bibitem[\protect\citeauthoryear{Fragile, Wilson  \& Rodriguez}{Fragile
  et~al.}{2012}]{Fragile2012a}
Fragile P.~C.,  Wilson J.,   Rodriguez M.,  2012, \mn@doi [\mnras]
  {10.1111/j.1365-2966.2012.21222.x}, 424, 524

\bibitem[\protect\citeauthoryear{Gressel}{Gressel}{2010}]{Gressel2010}
Gressel O.,  2010, \mn@doi [\mnras] {10.1111/j.1365-2966.2010.16440.x}, 405, 41

\bibitem[\protect\citeauthoryear{Gressel \& Pessah}{Gressel \&
  Pessah}{2015}]{Gressel2015}
Gressel O.,  Pessah M.~E.,  2015, \mn@doi [\apj] {10.1088/0004-637X/810/1/59},
  810, 59

\bibitem[\protect\citeauthoryear{Guilet \& Ogilvie}{Guilet \&
  Ogilvie}{2012}]{Guilet2012}
Guilet J.,  Ogilvie G.~I.,  2012, \mn@doi [\mnras]
  {10.1111/j.1365-2966.2012.21361.x}, 424, 2097

\bibitem[\protect\citeauthoryear{Guilet \& Ogilvie}{Guilet \&
  Ogilvie}{2013}]{Guilet2013}
Guilet J.,  Ogilvie G.~I.,  2013, \mn@doi [\mnras] {10.1093/mnras/sts551}, 430,
  822

\bibitem[\protect\citeauthoryear{Hawley, Gammie  \& Balbus}{Hawley
  et~al.}{1996}]{Hawley1996a}
Hawley J.~F.,  Gammie C.~F.,   Balbus S.~A.,  1996, \mn@doi [\apj]
  {10.1086/177356}, 464, 690

\bibitem[\protect\citeauthoryear{Hawley, Richers, Guan  \& Krolik}{Hawley
  et~al.}{2013}]{Hawley2013}
Hawley J.~F.,  Richers S.~A.,  Guan X.,   Krolik J.~H.,  2013, \mn@doi [\apj]
  {10.1088/0004-637X/772/2/102}, 772, 102

\bibitem[\protect\citeauthoryear{Hogg \& Reynolds}{Hogg \&
  Reynolds}{2018}]{Hogg2018a}
Hogg J.~D.,  Reynolds C.~S.,  2018, \mn@doi [\apj] {10.3847/1538-4357/aac439},
  861, 24

\bibitem[\protect\citeauthoryear{Igumenshchev, Narayan  \&
  Abramowicz}{Igumenshchev et~al.}{2003}]{Igumenshchev2003}
Igumenshchev I.~V.,  Narayan R.,   Abramowicz M.~A.,  2003, \mn@doi [\apj]
  {10.1086/375769}, 592, 1042

\bibitem[\protect\citeauthoryear{Jiang, Stone  \& Davis}{Jiang
  et~al.}{2014}]{Jiang2014b}
Jiang Y.-F.,  Stone J.~M.,   Davis S.~W.,  2014, \mn@doi [\apj]
  {10.1088/0004-637X/784/2/169}, 784, 169

\bibitem[\protect\citeauthoryear{Johansen \& Levin}{Johansen \&
  Levin}{2008}]{Johansen2008a}
Johansen A.,  Levin Y.,  2008, \mn@doi [\aap] {10.1051/0004-6361:200810385},
  490, 501

\bibitem[\protect\citeauthoryear{{Krause} \& {R{\"a}dler}}{{Krause} \&
  {R{\"a}dler}}{1980}]{KR80}
{Krause} F.,  {R{\"a}dler} K.~H.,  1980, {Mean-Field Magnetohydrodynamics and
  Dynamo Theory}.
Pergamon Press (also Akademie-Verlag: Berlin), Oxford

\bibitem[\protect\citeauthoryear{Liska, Tchekhovskoy  \& Quataert}{Liska
  et~al.}{2018}]{Liska2018a}
Liska M. T.~P.,  Tchekhovskoy A.,   Quataert E.,  2018, arXiv e-prints

\bibitem[\protect\citeauthoryear{Lubow, Papaloizou  \& Pringle}{Lubow
  et~al.}{1994}]{Lubow1994}
Lubow S.~H.,  Papaloizou J. C.~B.,   Pringle J.~E.,  1994, \mn@doi [\mnras]
  {10.1093/mnras/267.2.235}, 267, 235

\bibitem[\protect\citeauthoryear{Mandel}{Mandel}{1982}]{mandel_82}
Mandel J.,  1982, \mn@doi [The American Statistician]
  {10.1080/00031305.1982.10482771}, 36, 15

\bibitem[\protect\citeauthoryear{McKinney, Tchekhovskoy  \& Blandford}{McKinney
  et~al.}{2012}]{McKinney2012}
McKinney J.~C.,  Tchekhovskoy A.,   Blandford R.~D.,  2012, \mn@doi [\mnras]
  {10.1111/j.1365-2966.2012.21074.x}, 423, 3083

\bibitem[\protect\citeauthoryear{Meier}{Meier}{2005}]{Meier2005c}
Meier D.~L.,  2005, \mn@doi [\apss] {10.1007/s10509-005-1184-9}, 300, 55

\bibitem[\protect\citeauthoryear{{Mignone}, {Bodo}, {Massaglia}, {Matsakos},
  {Tesileanu}, {Zanni}  \& {Ferrari}}{{Mignone} et~al.}{2007}]{Mignone2007}
{Mignone} A.,  {Bodo} G.,  {Massaglia} S.,  {Matsakos} T.,  {Tesileanu} O.,
  {Zanni} C.,   {Ferrari} A.,  2007, \mn@doi [ApJS] {10.1086/513316}, \href
  {http://adsabs.harvard.edu/abs/2007ApJS..170..228M} {170, 228}

\bibitem[\protect\citeauthoryear{Mishra, Begelman, Armitage  \& Simon}{Mishra
  et~al.}{2019}]{Mishra2019a}
Mishra B.,  Begelman M.~C.,  Armitage P.~J.,   Simon J.~B.,  2019, arXiv
  e-prints

\bibitem[\protect\citeauthoryear{{Moffatt}}{{Moffatt}}{1978}]{Mof78}
{Moffatt} H.~K.,  1978, {Magnetic Field Generation in Electrically Conducting
  Fluids}.
Cambridge Univ.\ Press, Cambridge

\bibitem[\protect\citeauthoryear{Narayan \& Yi}{Narayan \&
  Yi}{1994}]{Narayan1994}
Narayan R.,  Yi I.,  1994, \mn@doi [\apjl] {10.1086/187381}, 428, L13

\bibitem[\protect\citeauthoryear{Narayan, Igumenshchev  \& Abramowicz}{Narayan
  et~al.}{2000}]{Narayan2000}
Narayan R.,  Igumenshchev I.~V.,   Abramowicz M.~A.,  2000, \mn@doi [\apj]
  {10.1086/309268}, 539, 798

\bibitem[\protect\citeauthoryear{Narayan, Igumenshchev  \& Abramowicz}{Narayan
  et~al.}{2003}]{Narayan2003}
Narayan R.,  Igumenshchev I.~V.,   Abramowicz M.~A.,  2003, \mn@doi [\pasj]
  {10.1093/pasj/55.6.L69}, 55, L69

\bibitem[\protect\citeauthoryear{Nauman \& Blackman}{Nauman \&
  Blackman}{2015}]{Nauman2015}
Nauman F.,  Blackman E.~G.,  2015, \mn@doi [\mnras] {10.1093/mnras/stu2226},
  446, 2102

\bibitem[\protect\citeauthoryear{Nemmen, Storchi-Bergmann  \& Eracleous}{Nemmen
  et~al.}{2014}]{Nemmen2014}
Nemmen R.~S.,  Storchi-Bergmann T.,   Eracleous M.,  2014, \mn@doi [\mnras]
  {10.1093/mnras/stt2388}, 438, 2804

\bibitem[\protect\citeauthoryear{{Paczy{\'n}sky} \& {Wiita}}{{Paczy{\'n}sky} \&
  {Wiita}}{1980}]{Paczynsky1980}
{Paczy{\'n}sky} B.,  {Wiita} P.~J.,  1980, \aap, \href
  {http://adsabs.harvard.edu/abs/1980A%26A....88...23P} {88, 23}

\bibitem[\protect\citeauthoryear{Papaloizou \& Pringle}{Papaloizou \&
  Pringle}{1984}]{Papaloizou1984}
Papaloizou J. C.~B.,  Pringle J.~E.,  1984, \mn@doi [\mnras]
  {10.1093/mnras/208.4.721}, 208, 721

\bibitem[\protect\citeauthoryear{Parkin \& Bicknell}{Parkin \&
  Bicknell}{2013}]{Parkin2013a}
Parkin E.~R.,  Bicknell G.~V.,  2013, \mn@doi [\mnras] {10.1093/mnras/stt1450},
  435, 2281

\bibitem[\protect\citeauthoryear{Penna, Narayan  \& S{\c a}dowski}{Penna
  et~al.}{2013}]{Penna2013}
Penna R.~F.,  Narayan R.,   S{\c a}dowski A.,  2013, \mn@doi [\mnras]
  {10.1093/mnras/stt1860}, 436, 3741

\bibitem[\protect\citeauthoryear{{Press}, {Teukolsky}, {Vetterling}  \&
  {Flannery}}{{Press} et~al.}{1992}]{recipies}
{Press} W.~H.,  {Teukolsky} S.~A.,  {Vetterling} W.~T.,   {Flannery} B.~P.,
  1992, {Numerical recipes in C. The art of scientific computing}

\bibitem[\protect\citeauthoryear{Quataert \& Gruzinov}{Quataert \&
  Gruzinov}{2000}]{Quataert2000}
Quataert E.,  Gruzinov A.,  2000, \mn@doi [\apj] {10.1086/309267}, 539, 809

\bibitem[\protect\citeauthoryear{{Racine}, {Charbonneau}, {Ghizaru}, {Bouchat}
  \& {Smolarkiewicz}}{{Racine} et~al.}{2011}]{racin_11}
{Racine} {\'E}.,  {Charbonneau} P.,  {Ghizaru} M.,  {Bouchat} A.,
  {Smolarkiewicz} P.~K.,  2011, \mn@doi [\apj] {10.1088/0004-637X/735/1/46},
  \href {https://ui.adsabs.harvard.edu/abs/2011ApJ...735...46R} {735, 46}

\bibitem[\protect\citeauthoryear{{R{\"a}dler}}{{R{\"a}dler}}{1969}]{Radler}
{R{\"a}dler} K.~H.,  1969, Veroeffentlichungen der Geod. Geophys, \href
  {https://ui.adsabs.harvard.edu/abs/1969VeGG...13..131R} {13, 131}

\bibitem[\protect\citeauthoryear{S{\c a}dowski, Narayan, Tchekhovskoy, Abarca,
  Zhu  \& McKinney}{S{\c a}dowski et~al.}{2015}]{Scadowski2015}
S{\c a}dowski A.,  Narayan R.,  Tchekhovskoy A.,  Abarca D.,  Zhu Y.,
  McKinney J.~C.,  2015, \mn@doi [\mnras] {10.1093/mnras/stu2387}, 447, 49

\bibitem[\protect\citeauthoryear{Schrinner, Rädler, Schmitt, Rheinhardt  \&
  Christensen}{Schrinner et~al.}{2007}]{sch_07}
Schrinner M.,  Rädler K.-H.,  Schmitt D.,  Rheinhardt M.,   Christensen U.~R.,
   2007, \mn@doi [Geophysical \& Astrophysical Fluid Dynamics]
  {10.1080/03091920701345707}, 101, 81

\bibitem[\protect\citeauthoryear{{Simard}, {Charbonneau}  \&
  {Dub{\'e}}}{{Simard} et~al.}{2016}]{simard_16}
{Simard} C.,  {Charbonneau} P.,   {Dub{\'e}} C.,  2016, \mn@doi [Advances in
  Space Research] {10.1016/j.asr.2016.03.041}, \href
  {https://ui.adsabs.harvard.edu/abs/2016AdSpR..58.1522S} {58, 1522}

\bibitem[\protect\citeauthoryear{{Stepanovs}, {Fendt}  \&
  {Sheikhnezami}}{{Stepanovs} et~al.}{2014}]{Stepanovs2014}
{Stepanovs} D.,  {Fendt} C.,   {Sheikhnezami} S.,  2014, \mn@doi [\apj]
  {10.1088/0004-637X/796/1/29}, \href
  {https://ui.adsabs.harvard.edu/abs/2014ApJ...796...29S} {796, 29}

\bibitem[\protect\citeauthoryear{{Sur}, {Brandenburg}  \& {Subramanian}}{{Sur}
  et~al.}{2008}]{Sur2008}
{Sur} S.,  {Brandenburg} A.,   {Subramanian} K.,  2008, \mn@doi [\mnras]
  {10.1111/j.1745-3933.2008.00423.x}, \href
  {https://ui.adsabs.harvard.edu/abs/2008MNRAS.385L..15S} {385, L15}

\bibitem[\protect\citeauthoryear{Suzuki \& Inutsuka}{Suzuki \&
  Inutsuka}{2014}]{Suzuki2014}
Suzuki T.~K.,  Inutsuka S.-i.,  2014, \mn@doi [\apj]
  {10.1088/0004-637X/784/2/121}, 784, 121

\bibitem[\protect\citeauthoryear{Tchekhovskoy, Narayan  \&
  McKinney}{Tchekhovskoy et~al.}{2011}]{Tchekhovskoy2011}
Tchekhovskoy A.,  Narayan R.,   McKinney J.~C.,  2011, \mn@doi [\mnras]
  {10.1111/j.1745-3933.2011.01147.x}, 418, L79

\bibitem[\protect\citeauthoryear{{Tomei}, {Del Zanna}, {Bugli}  \&
  {Bucciantini}}{{Tomei} et~al.}{2020}]{Tomei2020}
{Tomei} N.,  {Del Zanna} L.,  {Bugli} M.,   {Bucciantini} N.,  2020, \mn@doi
  [\mnras] {10.1093/mnras/stz3146}, \href
  {https://ui.adsabs.harvard.edu/abs/2020MNRAS.491.2346T} {491, 2346}

\bibitem[\protect\citeauthoryear{Velikhov}{Velikhov}{1959}]{Velikhov1959}
Velikhov E.,  1959, Sov. Phys. JETP, 36, 995

\bibitem[\protect\citeauthoryear{{Yuan} \& {Narayan}}{{Yuan} \&
  {Narayan}}{2014}]{Yuan2014}
{Yuan} F.,  {Narayan} R.,  2014, \mn@doi [\araa]
  {10.1146/annurev-astro-082812-141003}, \href
  {https://ui.adsabs.harvard.edu/abs/2014ARA&A..52..529Y} {52, 529}

\bibitem[\protect\citeauthoryear{{Yuan}, {Bu}  \& {Wu}}{{Yuan}
  et~al.}{2012}]{Yuan2012}
{Yuan} F.,  {Bu} D.,   {Wu} M.,  2012, \mn@doi [\apj]
  {10.1088/0004-637X/761/2/130}, \href
  {https://ui.adsabs.harvard.edu/abs/2012ApJ...761..130Y} {761, 130}

\bibitem[\protect\citeauthoryear{{Zhu} \& {Stone}}{{Zhu} \&
  {Stone}}{2018}]{Zhu2018}
{Zhu} Z.,  {Stone} J.~M.,  2018, \mn@doi [\apj] {10.3847/1538-4357/aaafc9},
  \href {https://ui.adsabs.harvard.edu/abs/2018ApJ...857...34Z} {857, 34}

\makeatother
\end{thebibliography}
\label{lastpage}

\appendix
\section{Complete Set of Dynamo Coefficients}
\label{app_all_coefficients}
Without neglecting the contribution of $\mathbf{b}$ tensor in the 
expansion of equation \ref{eq:mean_emf_soca}, $i^{th}$ component of EMF 
can be written as 
\bea
\bar{\mathcal{E}}_i = \tilde{a}_{ij} \bar{B}_j + \tilde{b}_{ijr} \frac{\partial \bar{B}_j}{\partial r}
+ \frac{\tilde{b}_{ij\theta}}{r} \frac{\partial \bar{B}_j}{\partial \theta},
\eea
and the components of pseudo tensors $\tilde{a}_{ij}$ and 
$\tilde{b}_{ij}$ can then be used to express the following 
complete set of dynamo coefficients including the diffusive 
terms, that depend explicitly on the components of $\mathbf{
b}$,

\bea
\label{eq:alphas_complete}
\alpha_{rr}             &=& \tilde{a}_{rr} - \frac{\tilde{b}_{r\theta\theta}}{r},\nonumber\\
\alpha_{\theta \theta}  &=& \tilde{a}_{\theta \theta} 
                            +   \frac{\tilde{b}_{\theta r\theta}}{r},\nonumber\\
\alpha_{\phi\phi}       &=&  \tilde{a}_{\phi \phi},\nonumber\\
\alpha_{r\theta}        =& \alpha_{\theta r} = &\frac{1}{2}\left(
                                \tilde{a}_{r \theta} + \tilde{a}_{\theta r} 
                            +   \frac{\tilde{b}_{rr\theta}}{r}
                            -   \frac{\tilde{b}_{\theta\theta\theta}}{r}
                                \right),\nonumber\\
\alpha_{r\phi}          =& \alpha_{\phi r} =& \frac{1}{2}
                                \left(\tilde{a}_{r \phi} + \tilde{a}_{\phi r} 
                            -   \frac{\tilde{b}_{\phi rr}}{r} 
                                \right),\nonumber\\
\alpha_{\theta\phi}     =& \alpha_{\phi\theta} = &\frac{1}{2}
                                \left(\tilde{a}_{\theta \phi} + \tilde{a}_{\phi \theta} 
                            +   \frac{\tilde{b}_{\phi r \theta}}{r} 
                                \right).
\eea

\bea
\label{eq:gammas_complete}
\gamma_r        &=&   \frac{1}{2}
                    \left(
                        \tilde{a}_{\theta\phi}
                    -   \tilde{a}_{\phi\theta}
                    -   \frac{\tilde{b}_{\phi r\theta}}{r}
                    \right),\nonumber\\    
\gamma_\theta   &=&   \frac{1}{2}
                    \left(
                        \tilde{a}_{\phi r}
                    -   \tilde{a}_{r \phi}
                    -   \frac{\tilde{b}_{\phi\theta\theta}}{r}                    
                    \right),\nonumber\\   
\gamma_\phi     &= &  \frac{1}{2}
                    \left(
                        \tilde{a}_{r\theta}
                    +   \frac{\tilde{b}_{rr\theta}}{r}
                    -    \tilde{a}_{\theta r}
                    +   \frac{\tilde{b}_{\theta\theta\theta}}{r}                    
                    \right).   
\eea

\bea
\label{eq:betas_complete}
\eta_{rr}              &=&   -   \frac{1}{2} \tilde{b}_{r\phi\theta},\nonumber\\
\eta_{\theta\theta}    &=& \frac{1}{2} \tilde{b}_{\theta\phi r},\nonumber\\
\eta_{\theta\theta}    &=& \frac{1}{2} \left(\tilde{b}_{\phi r \theta} 
                            -   \tilde{b}_{\phi \theta r} \right),\nonumber\\
\eta_{r\theta}         =& \eta_{\theta r}  = &\frac{1}{4}\left(
                                \tilde{b}_{r \phi r} 
                            -   \tilde{b}_{\theta \phi \theta}
                                \right),\nonumber\\
\eta_{r\phi}           =& \eta_{\phi r}    = &\frac{1}{4}\left(
                                \tilde{b}_{r r \theta} 
                            -   \tilde{b}_{\phi \phi \theta}
                            -   \tilde{b}_{r \theta r}
                                \right),\nonumber\\ 
\eta_{\theta\phi}      =& \eta_{\phi \theta}    =& \frac{1}{4}\left(
                                \tilde{b}_{\theta r \theta} 
                            +   \tilde{b}_{\phi \phi r}
                            -   \tilde{b}_{\theta \theta r}
                                \right).
\eea

\bea
\label{eq:deltas_complete}
\Delta_r        &=&   \frac{1}{4}
                    \left(
                        \tilde{b}_{\theta \theta r}
                    -   \tilde{b}_{\theta r \theta}
                    +   \tilde{b}_{\phi \phi r}
                    \right),\nonumber\\    
\Delta_\theta   &=&   \frac{1}{4}
                    \left(
                        \tilde{b}_{r r \theta}
                    -   \tilde{b}_{r \theta r}
                    +   \tilde{b}_{\phi \phi \theta}
                    \right),\nonumber\\   
\Delta_\phi     &= &  -\frac{1}{4}
                    \left(
                        \tilde{b}_{r \phi r}
                    +   \tilde{b}_{\theta \phi \theta}
                    \right).   
\eea

\bea
\label{eq:kappas_complete}
\kappa_{irr}            &=& - \tilde{b}_{irr} ,\nonumber\\
\kappa_{i\theta\theta}  &=& - \tilde{b}_{i\theta\theta} ,\nonumber\\
\kappa_{i\phi\phi}      &=& 0 ,\nonumber\\
\kappa_{ir\theta}       =& \kappa_{i\theta r} =& 
                            - \frac{1}{2}\left(
                              \tilde{b}_{ir\theta} + \tilde{b}_{i\theta r}
                              \right),\nonumber\\
\kappa_{ir\phi}         =& \kappa_{i\phi r} =& 
                            - \frac{1}{2}\tilde{b}_{i\phi r},\nonumber\\
\kappa_{i\theta\phi}    =& \kappa_{i\phi\theta} =& 
                           -  \frac{1}{2} \tilde{b}_{i\phi\theta},
\eea

\end{document}